\documentclass [longauth]{aa}
\usepackage[]{natbib}
\bibpunct{(}{)}{;}{a}{}{,}
\usepackage{subfig}
\usepackage{graphicx}
\usepackage{epsf}
\usepackage{rotating}
\usepackage{graphics}
\usepackage{txfonts}
\usepackage{amssymb}
\newcommand{\corot}{\emph{CoRoT}}

\newcommand{\corotfour}{{CoRoT-Exo-4b}}
\newcommand{\corotp}{{CoRoT-Exo-3b}}
\newcommand{\corots}{{CoRoT-Exo-3}}

\newcommand{\sophie}{\emph{SOPHIE}}
\newcommand{\tls}{\emph{TLS}}
\newcommand{\coralie}{\emph{CORALIE}}
\newcommand{\harps}{\emph{HARPS}}

\newcommand{\uves}{\emph{UVES}}

\newcommand{\MJ}{M$_{Jup}$}
\newcommand{\RJ}{R$_{Jup}$}
\newcommand{\sn}{S/N}
\newcommand{\kms}{km\,s$^{-1}$}
\newcommand{\ms}{m\,s$^{-1}$}
\newcommand{\gc}{g\,cm$^{-3}$}
\newcommand{\exodat}{\emph{Exo-Dat}}

\newcommand{\mass}{\emph{2-MASS}}

\newcommand{\ucac}{\emph{UCAC2}}

\newcommand{\vsini}{$v$\,sin\,$i$}   
\newcommand{\vrad}{$v_{\rm rad}$} 

\newcommand{\teff}{T$_{\rm eff}$}
\newcommand{\logg}{log {\it g}}
\newcommand{\met}{[M/H]}
\newcommand{\mr}{$M_\star^{1/3}/R_\star$}

\begin{document}
\title{Transiting exoplanets from the \corot\ space mission \thanks{The \corot\ space mission, launched on December 27th 2006, has been developed and is operating by CNES, with the contribution of Austria, Belgium, Brasil, ESA, Germany and Spain. The first \corot\ data will be available to the public in February 2009 from the \corot\ archive: http://idoc-corot.ias.u-psud.fr/}}

\subtitle{VI.  \corotp: The first secure inhabitant of the brown-dwarf desert} 
   \author{Deleuil, M.\inst{1}
      \fnmsep
      \and
      Deeg, H.J. \inst{2}
      \and
      Alonso, R. \inst{1}
      \and
      Bouchy, F. \inst{3}
      \and 
      Rouan, D. \inst{4}
      \and 
      Auvergne, M. \inst{4}
      \and 
      Baglin, A. \inst{4}
\and Aigrain, S. \inst{5}
\and Almenara, J.M. \inst{2}
\and Barbieri, M. \inst{1}
\and Barge, P. \inst{1}
\and Bruntt, H. \inst{6}
\and Bord\'e, P. \inst{7}
\and Collier Cameron, A. \inst{8}
\and Csizmadia, Sz. \inst{9}
\and De la Reza, R. \inst{10}
\and Dvorak, R. \inst{11}
\and Erikson, A. \inst{9}
\and Fridlund, M. \inst{12}
\and Gandolfi, D. \inst{13}
\and Gillon, M. \inst{14}
\and Guenther, E. \inst{13}
\and Guillot, T. \inst{15}
\and Hatzes, A. \inst{13}
\and H\'ebrard, G. \inst{3}
\and Jorda, L. \inst{1}
\and Lammer, H. \inst{16}
\and L\'eger, A. \inst{7}
\and Llebaria, A. \inst{1}
\and Loeillet, B. \inst{1,3}
\and Mayor, M. \inst{14}
\and Mazeh, T. \inst{17}
\and Moutou, C. \inst{1}
\and Ollivier, M. \inst{7}
\and P\"atzold, M. \inst{18}
\and Pont, F. \inst{5}
\and Queloz, D. \inst{14}
\and Rauer, H. \inst{9,19}
\and Schneider, J. \inst{20}
\and Shporer, A. \inst{17}
\and Wuchterl, G. \inst{13}
\and Zucker, S. \inst{17}
      }
      
   \offprints{M. Deleuil,
     \email magali.deleuil@oamp.fr}

   \institute{Laboratoire d'Astrophysique de Marseille (UMR 6110), Technop™le de Marseille-Etoile, F-13388 Marseille cedex 13 (France)
          \email{magali.deleuil@oamp.fr}
     \and
         Instituto de Astrof\'\i sica de Canarias, C. Via Lactea S/N, E-38200 La Laguna (Spain)
     \and
       Institut d'Astrophysique de Paris, UMR7095 CNRS, Universit\'e Pierre \& Marie Curie, 98bis Bd Arago, 75014 Paris, France
\and 
LESIA, CNRS UMR 8109, Observatoire de Paris, 5 place J. Janssen, 92195 Meudon, France
\and
School of Physics, University of Exeter, Stocker Road, Exeter EX4 4QL, United Kingdom
\and
School of Physics A28, University of Sydney, Australia
\and
IAS, Universit\'e Paris XI, 91405 Orsay, France
\and
School of Physics and Astronomy, University of St Andrews, United Kingdom
\and
Institute of Planetary Research, DLR, Rutherfordstr. 2, 12489 Berlin, Germany
\and
Observat\'orio Nacional, Rio de Janeiro, RJ, Brazil
\and
Institute for Astronomy, University of Vienna, T\"urkenschanzstrasse 17, 1180 Vienna, Austria
\and
Research and Scientific Support Department, European Space Agency, ESTEC, 2200 Noordwijk, The Netherlands 
\and
Th\"uringer Landessternwarte Tautenburg, Sternwarte 5, 07778 Tautenburg, Germany
\and
Observatoire de Gen\`eve, Universit\'e de Gen\`eve, 51 Ch. des Maillettes, 1290 Sauverny, Switzerland
\and
Observatoire de la C\^ote d'Azur, Laboratoire Cassiop\'ee, CNRS UMR 6202, BP 4229, 06304 Nice Cedex 4, France
\and
Space Research Institute, Austrian Academy of Sciences, Schmiedlstrasse 6, 8042 Graz, Austria
\and
School of Physics and Astronomy, R. and B. Sackler Faculty of Exact Sciences, Tel Aviv University, Tel Aviv 69978, Israel
\and 
Rheinisches Institut f\"ur Umweltforschung, Universit\"at zu K\"oln, Abt. Planetenforschung, Aachener Str. 209, 50931 K\"oln, Germany
\and
Center for Astronomy and Astrophysics, TU Berlin, Hardenbergstr. 36, D-10623 Berlin, Germany  
\and
LUTH, Observatoire de Paris-Meudon, 5 place J. Janssen, 92195 Meudon, France
}      

   \date{}
  \abstract
  {The \corot\ space mission routinely provides high-precision photometric measurements of thousands of stars that have been continuously observed for months. 
   } 
{The discovery and characterization of the first very massive transiting 
planetary companion with a short orbital period is reported.}
{A series of 34 transits was detected in the \corot\ light curve of an F3V star, observed from May to October 2007 for 152 days. The radius was accurately determined and the mass derived for this new transiting, thanks to the combined analysis of the light curve and complementary ground-based observations: high-precision radial-velocity measurements, on-off photometry, and high signal-to-noise spectroscopic observations.}
{\corotp\  has a radius of 1.01 $\pm$ 0.07 \RJ\ and transits around its F3-type primary every 4.26 days in a synchronous orbit. Its mass of 21.66 $\pm$ 1.0 \MJ, density of 26.4 $\pm 5.6$ $g\,cm^{-3}$, and surface gravity of $\log g = 4.72$ clearly distinguish it from the regular close-in planet population, making it the most intriguing transiting substellar object discovered so far. }
{With the current data, the nature of \corotp\ is ambiguous, as it could either be a low-mass brown-dwarf or 
a member of a new class of ``superplanets". Its discovery may help constrain the evolution of close-in planets and brown-dwarfs better. Finally, \corotp\ confirms the trend that massive transiting giant planets (M $\ge$ 4 \MJ) are found preferentially around more massive stars than the Sun. }
{}
 
  \authorrunning{Deleuil M. et al.}
  \titlerunning{\corotp: first transiting close-in substellar object}
  \maketitle
  
%

\section{Introduction}
Massive close-in planets are nowadays the
most accessible population of extrasolar planets, and they are extensively being
studied with both radial velocity and transit surveys. To date, the more than forty extrasolar transiting planets with known mass and radius indeed belong to this population. However, the ever increasing number 
of discovered new members of this group widens the range of 
their properties and
challenges our understanding of their formation and structure. 

As demonstrated by its first results \citep{2008A&A...482L..17B,2008A&A...482L..21A}, the space mission \corot\ is particularly well-suited to making significant breakthroughs in our knowledge of this population of planets in short orbital periods. The instrument is performing wide-field, relative stellar photometry at ultra-high precision. It can monitor up to 12~000 stars simultaneously per observing run, over a temporal span up to 150 days of nearly continuous observations. It is thus sensitive to detecting
planets with orbital periods less than 75 days.  One advantage of
CoRoT's performance is that it  nicely matches that of
ground-based radial velocity facilities. Combining the ultra-precise
stellar photometry with precise radial velocity observations enable us to 
fully investigate the nature of the discovered objects.

We report in this paper the discovery by \corot\ of the smallest 
close-in transiting substellar object detected so far by photometry. The analysis of the high-quality \corot\ 
light curve, combined with follow-up observations, allow us to fully characterize this new intriguing object. It orbits in  4.25680 days an F-type dwarf with solar metallicity. 
From the derived mass we know that this object is located in the
so-called brown-dwarf desert. This is the gap in the mass function
that separates stellar and planetary objects.

\section{\corot observations}
\subsection{Light curve}
{\corots} was observed during the first long observing run of \corot\ ({\sl LRc01}) which took place from May 26th to October 25th 2007. This stellar field is centered at $\alpha = 19^{\rm h} 23^{\rm m}$, $\delta$ = $\delta = +00^{\circ} 27^{\prime}$ ~in a direction close to the Galaxy center. With a magnitude of R=13.1 (Table~\ref{StarID}), the host star is among the 
the brightest \corot\ targets, which are typically 
in the range 11 to 16 in $R$. It is one of the targets identified as a planetary candidate in the so-called {\it ``alarm mode''} (\citealt{Quentin06}; Surace et al, 2008). Following this detection, on August 3, 2007 the time sampling of the light curve was switched from 512~sec, corresponding to the co-addition on-board of 16 elementary exposures, to the nominal 32~sec.
As a whole, the light curve consists of 236999 measurements with a 
time sampling of 512~sec for the first two months, and then 32~sec sampling
until the end of the observational run. The data presented in this paper were reduced with the very first version of the \corot\ calibration pipeline (Auvergne et al., {\sl in preparation}). This current version of the pipeline includes (i) the correction for  the CCD zero offset and gain, (ii) the correction for  the background contribution, which is done using reference photometric windows, 
and (iii) the correction for the satellite orbital effects.
In addition, the pipeline flags outliers due to impacts of charged particles. The  flux of highly energetic particles increases dramatically 
during the crossing of the South Atlantic Anomaly (SAA) and, despite the shielding of the focal plane, makes photometric measurements impossible. The corresponding portions of the light curve were simply removed from the final light curve. This results in a duty cycle of about 88\%. 
\begin{figure*}[ht]
\begin{center} 
 \includegraphics[width=18cm,height=8cm]{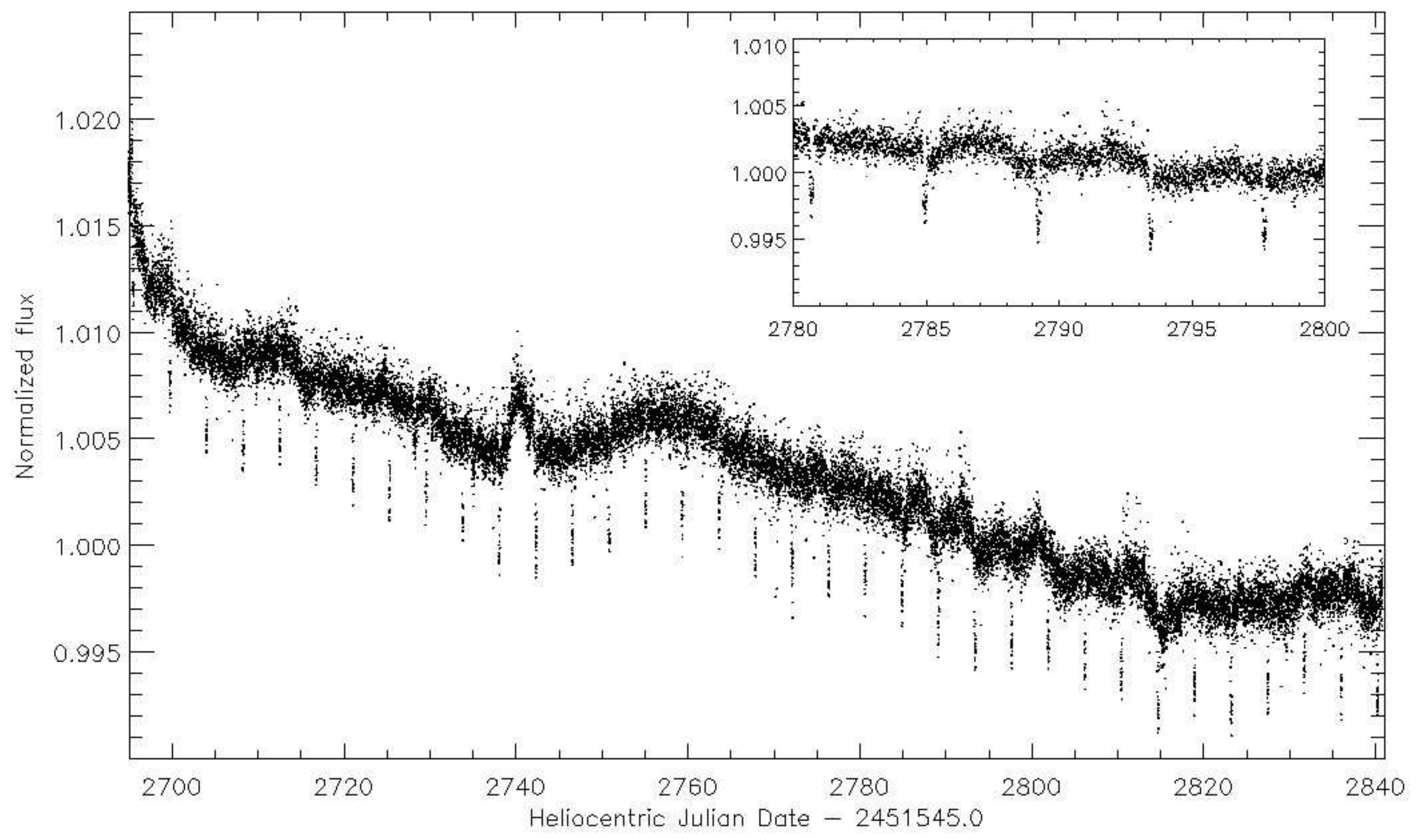}
\end{center}
\caption{The 152-days long \corotp\ reduced light curve, binned at a constant time sampling of 512~sec and normalized. The low activity level of the star is illustrated in the upper figure which displays a zoomed section of the light curve.}
\label{LCTotal}
\end{figure*}

\begin{table}[h]
\caption{\label{StarID} 
\corots\ IDs, coordinates and magnitudes with errors.}
\begin{center}{
\begin{tabular}{rcc }
\hline
\hline
 \corot\ ID & \multicolumn{2}{c}{ 101368192} \\
USNO-A2         & \multicolumn{2}{c}{0900-15209129}\\ 
2MASS            &  \multicolumn{2}{c}{19281326+0007185}                      \\
GSC2.3 & \multicolumn{2}{c}{N1MO000645}\\ 
RA (2000)       & \multicolumn{2}{c}{19:28:13.26  } \\
DEC (2000)     & \multicolumn{2}{c}{ 00:07:18.7 } \\
\hline
B-mag~$^{\mathrm{(a)}}$             &  14.199 & 0.04\\
V-mag~$^{\mathrm{(a)}}$             &  13.292 & 0.02\\
$r^\prime$-mag~$^{\mathrm{(a)}}$  &  13.099  & 0.01 \\
$i^\prime$-mag~$^{\mathrm{(a)}}$  & 12.676  & 0.01 \\
J ~$^{\mathrm{(b)}}$     & 11.936 & 0.028  \\
H~$^{\mathrm{(b)}}$      & 11.710 &  0.041  \\
K$_s$~$^{\mathrm{(b)}}$  & 11.618 & 0.033 \\
\hline
$\mu_\alpha$~$^{\mathrm{(c)}}$ & -10.8 & 5.6 \arcsec/yr  \\
$\mu_\delta$~$^{\mathrm{(c)}}$ & -10.8 &  5.6 \arcsec/yr  \\
\hline
\end{tabular}}
\begin{footnotesize}
\begin{list}{}{}
\item (a) Provided by \exodat, based on observations taken at the {\sl INT} telescope. 
\item (b) From \mass\ catalog.
\item (c) From \ucac\ catalog.
\end{list}
\end{footnotesize}
\end{center}
\end{table}

In a final step, we also corrected the light curve from the signatures of pixel defects, often called ``hot-pixels''. These hot-pixels are a direct 
consequence of high energy particle impacts onto the \corot\ CCD \citep[see][]{2008MNRAS.384.1337P} and cause discontinuities in the light curve at the time of their creation. Different kinds of discontinuities are observed: a sudden increase of the apparent flux of the light curve with an exponential decay or a step-like discontinuity. We corrected these temporal signatures, case by case, by an exponential function with a time constant or a simple step function. The resulting light curve is shown in Fig~\ref{LCTotal}. We note a gradual decrease of the signal of about 2\%, a behavior observed in all the \corot\ light curves and whose origin is ascribed to CCD aging. 
The \sn\ per 512~$\sec$, estimated on out of transit sections of the light curve, is $\simeq$1915.

When analyzing \corot\ data, special care should be taken to account for the possible contamination of the target's signal by nearby stars whose Point Spread Function (PSF) falls within the quite large aperture mask of \corot. We checked in the \exodat\ database (Deleuil et al., {\it submitted}) the vicinity of the target  (see Sect.~\ref{PhotomFU}). \exodat\ is the \corot/Exoplanet Scientific database which was used to build the Exoplanet input catalog and gathers all information on the target stars as well as their environment. Among all the stars whose flux could contribute to the light curve, the brightest one is 2.8 magnitudes fainter in the R-band and located at 5.6~\arcsec\ to the South. To evaluate the contribution of this and further nearby stars we followed the method described in \cite{2008A&A...482L..21A}, which gives us a value of about 8\%. To properly remove its contribution to the white light curve, we identified in \exodat\ a set of a dozen of stars with similar magnitude to the contaminant,  observed by \corot\ during the same run and on the same CCD. We checked that these stars were not affected by nearby contaminants so that their light curve could be used as reference.
We computed the median value of the light curves of these stars and found a 
fractional flux of 0.082 $\pm$ 0.007. This value was then subtracted from the light curve. 

\subsection{Transit parameters}
\begin{figure}
   \centering
    \includegraphics[width=9cm]{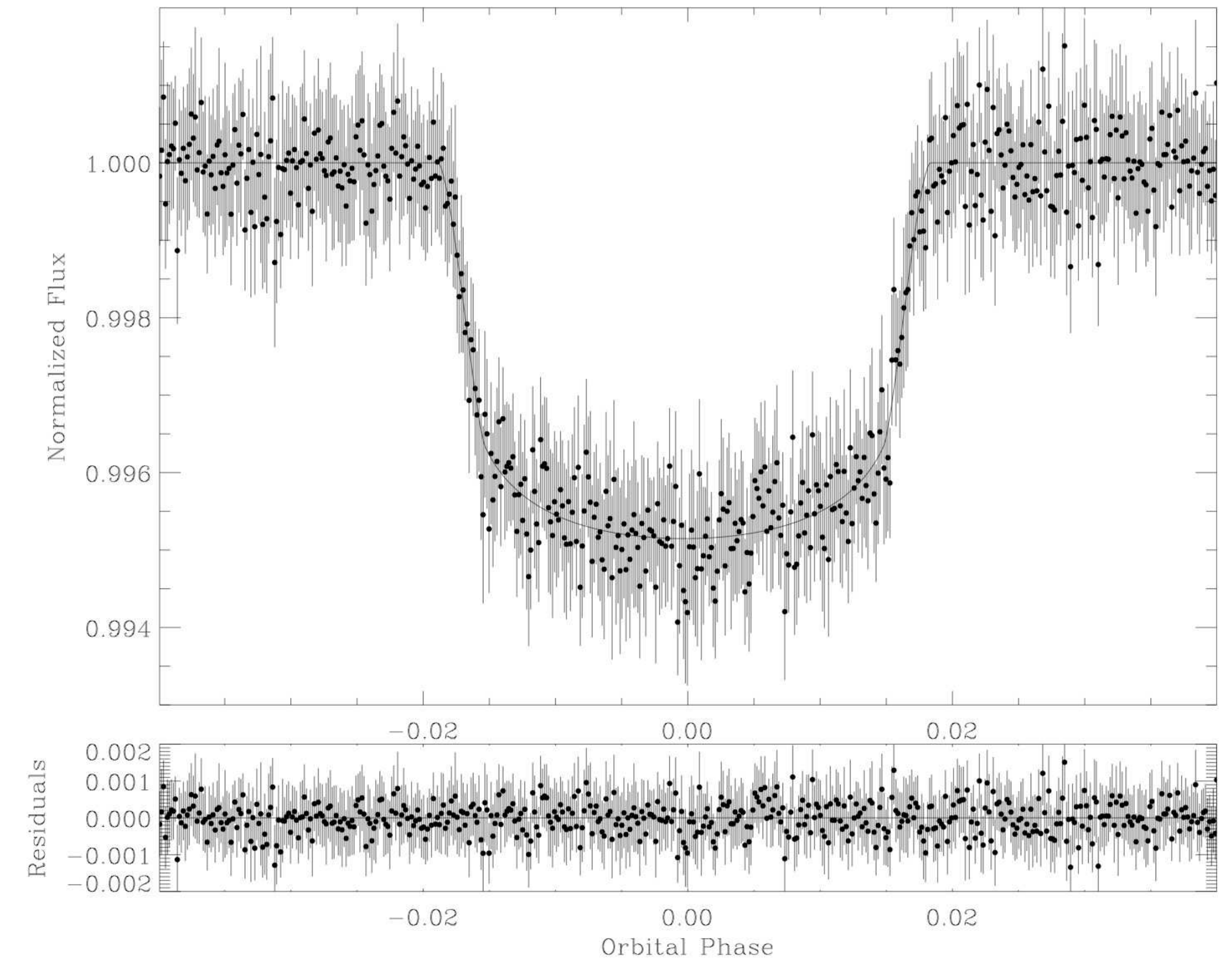}
   \caption{The phase-folded light curve of the \corotp\ transit with the best fit solution overplotted (solid line). The residuals of the fit are plotted at the bottom of the figure.  The rms error is 4.4094 10$^{-4}$ out of eclipse.}              
\label{TransitFit}
    \end{figure}

A total of 34 transits are visible in the light curve, 18 belonging to
the 32~sec sampling part (Fig~\ref{LCTotal}). In itself, the light curve does not exhibit strong photometric variations, indicating a non-active star. The transit was analyzed using the same methodology as presented in \cite{2008A&A...482L..21A}, though the very low activity level of the star facilitates the analysis. 
We simply recall here the main steps of the method. From the series of transits, both the orbital period and the transit epoch were derived by a trapezoidal fitting to all the transit centers. The light curve was phase-folded
to this  ephemeris after first
performing a local linear fit to the region centered around the transit in
order to correct for any local variation of the light curve over 
a range from $\pm$0.02 to $\pm$0.04 in phase. The transit light curve
shown in Fig~\ref{TransitFit} was binned by  1.5 10$^{-3}$ in phase 
and the error bar of each individual bin calculated as the dispersion of the points inside the bin, divided by the square root of the number of points per bin. The system parameters were then estimated by a $\chi^{2}$ fitting method, following the formalism of \cite{2006A&A...450.1231G}. Basically, the model is a 6-free parameters model: the transit center, the orbital phase at first contact
between planet and star, the ratio of the planet to star radii, 
$R_p/R_\star$, the orbital inclination $i$, and the two coefficients $u_+=u_a +u_b$ and $u_-=u_a-u_b$, where $u_a, u_b$ are the quadratic limb-darkening coefficients \citep{2003A&A...401..657C,2004A&A...428.1001C}. 
\begin{table}
\begin{center}{
\caption{\label {TransitParam} Transit parameters derived from the
\corot\ light curve analyses.}
\begin{tabular}{lll}
\hline
\hline
    & \multicolumn{2}{c}{best fit}  \\
    & \multicolumn{2}{c}{all parameters free}   \\
   Parameter      & Value   &  Uncertainty   \\
\hline
Period (day)  &  4.25680 & $\pm$ 0.000005      \\
T$_0$ (day) &  2454283.1383 & $\pm$ 0.0003        \\
$\chi^2$       &   1.020 &     \\
$\theta_1$   &   0.0185  &  $\pm$ 1.6 10$^{-4} $  \\ 
$k = R_p/R_\star$ &   0.0663  & $\pm$  9.0 10$^{-4}$   \\ 
$i$ (deg) &   85.9  & $\pm$ 0.8   \\
$u_{+}$ &  0.56   &  $\pm$ 0.05   \\
$u_{-}$ &  $-$0.10  &   $\pm$  0.07 \\ 
\mr\ &   0.71  &  $\pm$  0.04  \\
$a/R_\star$  &   7.8  &  $\pm$ 0.4    \\
$b~(R_\star)$~$^{\mathrm{(a)}}$  &   0.55  &  $\pm$ 0.08     \\
\hline
\end{tabular}}
\begin{footnotesize}
\begin{list}{}{}
\item (a) Impact parameter
\end{list}
\end{footnotesize}
\end{center}
\end{table}
For the transit fitting, we used the same method as fully described in \cite{2008A&A...482L..17B} and \cite{2008A&A...482L..21A}, with a bootstrap analysis to properly explore the parameter space. We found (Table~\ref{TransitParam}) that the best fit gives values for the two limb-darkening parameters with large error bars, especially for the $u_{-}$ parameter.  

The limb-darkening parameters issue has been already pointed out by various studies carried out using space-based or ground-based observations \citep{2007ApJ...655..564K,2008MNRAS.386.1644S}. In the case of \corots, as the broad \corot\ spectral band-pass is unique and not related to any of the photometric filters used for limb-darkening predictions, we decided not to fix the limb-darkening parameters in the transit fitting. We however checked that the values obtained with the best fit model were consistent with the expected theoretical values.
 
According to Claret (2004) for a given value of \teff\ and \logg, the $u_a$ and $u_b$ values change slightly from one filter to another. As the maximum of transmission of \corot\ is close to the V and R bands, 
we assess the range of predicted values for $u_a$ and $u_b$ respectively, in the $B$, $V$, $R$ and $i$ filters. This was done for different values of \teff\ and \logg\ which correspond to the physical properties and their uncertainties 
that we derived for \corots\ (see Sect.~\ref{SpectroParam}). As a result, this gives us a range of $u_a$ and $u_b$ predicted values, which simply translates into a range of possible values for both $u_{-}$ and $u_{+}$. We checked that our best fitted values are well within each of these ranges.  

It is worth noticing that we found another set of solutions with comparable $\chi^2$ value but at a higher inclination, $i \simeq 89^{\circ }$. These solutions however give fitted limb-darkening parameters which are even more inconsistent with theoretical ones. As reported by \cite{2001ApJ...552..699B}, for transiting planets at low impact parameters, there is a degeneracy in the orbital inclination and the limb darkening coefficients, both effects being difficult to disentangle at the level greater than a few 10$^{-4}$. We compared the two solutions, and checked at the level of precision we achieve, with a rms error of 4.4 10$^{-4}$ out of transit that they could not be distinguished.

\section{Follow-up observations} 

Ground-based follow-up campaigns of the planet candidates detected by 
the {\sl alarm mode} in 
the \corot\ light curves from the first long run started in July 2007, or
shortly after their detection. Such complementary observations are mandatory to identify the nature of the transiting bodies and to further characterize the secured planets. Radial velocity measurements as well as ground-based photometry observations were performed and confirmed the non-stellar nature of the transiting body. High resolution spectra of the host star were later acquired with the \uves\ spectrograph on the VLT. 
 
\subsection{Photometric follow-up observations}
\label{PhotomFU}
\begin{figure}
   \centering
 \includegraphics[width=6cm]{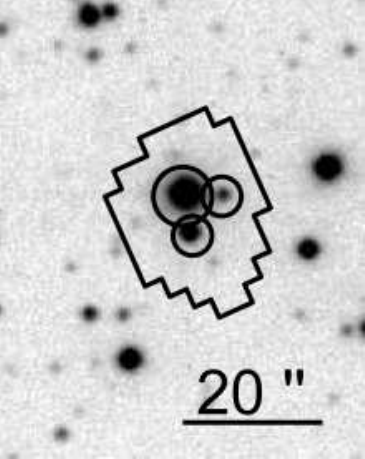}
   \caption{A CCD image of the field around \corots, observed
with the IAC80 telescope at a seeing with a FWHM of 1.4\arcsec. The
largest circle indicates the \corotp\ host
star, whereas the smaller circles below and to the right indicate contaminating stars that are 2.9~mag  and 4.9~mag fainter, at distances of 5.6\arcsec to the South and 5.3\arcsec to the East, respectively. The irregular shape is the photometric mask used by \corot. Please note that the East is to the right on the figure.}              
\label{fig:exo3field}
    \end{figure}

The relatively large PSF of \corot\ (of 20\arcsec x 6\arcsec \, with the longer axes approximately in N-S direction) and its large aperture masks raise the possibility that any of the
nearby fainter stars (see Fig.~\ref{fig:exo3field}) that contaminate
a target lightcurve might in fact be an eclipsing binary mimicking
a transit-like signal for the target. Indeed, around \corots\ there are two such \emph{potential} sources for false alarms, at distances of 5.3 and 5.6\arcsec, respectively.
As part of the ongoing ground-based
photometric follow-up of \corot\ candidates, whose motivation and techniques are described in more detail in
Deeg et al 2008 ({\it in preparation}), time-series imaging was
obtained during both on-transit and off-transit configurations on separate nights in July
and August 2007 with the IAC 80cm telescope at a spatial resolution of
 about 1.4\arcsec. A comparison of the
brightnesses of the transit-candidate as well as of the potential false alarm sources was
then performed between the timeseries taken at ``on'' and ``off''-transit configurations.
While this comparison could not reproduce Corot's observed brightness
variation of 5~$mmag$ on the candidate star, the {\it absence} of any sufficiently large brightness variation (of at least 0.06 and 0.4\, mag for the brighter and the fainter of the contaminants, respectively) could be
demonstrated. We can therefore deduce that \corot's transit-detection
arises from a small brightness variation on the target-star itself. 
Finally, a full transit on the target star was
observed at the 1-m {\sl WISE} telescope on 5 July 2008.  The shape of the target light curve was consistent with the \corot\ light curve, in both transit depth and duration. It nicely confirms the \corot's transit detection.

\subsection{Radial velocity measurements}

In the days following its detection, \corots\ was observed with the \sophie\ spectrograph \citep{2006tafp.conf..319B} on the 1.93-m OHP telescope (France). A first set of 5 spectra was recorded between July, 28 and August 23, 2007, in high efficiency mode (spectral resolution $R \simeq $ 40 000). Additional measurements were made more than 9 months after, in 2008, in order to investigate a possible drift due to a second companion. The spectra were extracted using the on-line \sophie\ reduction pipeline which allows us  immediately
to perform the radial velocity analysis  and thus to rapidly reject 
false positives. The radial velocities were measured by cross-correlation of the reduced spectra with a numerical G2 mask, constructed from the Sun spectrum atlas including up to 3645 lines. Nearly at the same dates, the star was observed  with the Coud\'e \'echelle spectrograph of the 2-m-Alfred Jensch telescope in Tautenburg, Germany (\tls). In three different nights, two spectra were recorded, each of 30~$\min$ exposure time.  
Using the so-called VIS grism and a slit width of 2~$\arcsec$, these spectra cover
the wavelength range from 470 to 740 nm at a resolution of $\lambda /
\Delta\lambda =33 000$, giving about 4 pixels per
resolution element.  Moonlight was removed by taking a spectrum of the
moon, scaled to the level measured along the 30~$\arcsec$
long slit and then subtracting it. For flat-fielding we used a
specially designed dome-flat utility, and for the wavelength
calibration a ThAr-lamp. The spectra were bias-subtracted,
flat-fielded, cosmic rays removed and extracted using standard IRAF
routines. After accounting for an instrumental shift using the
telluric lines, the radial velocity of the star was finally measured by
cross-correlating the spectrum with the radial velocity standard star
\object{HR 5777} for which we used a radial velocity of
$+49.12\pm0.06$ $kms^{-1}$ \citep{1993ApJ...413..349M}. For the
cross-correlation, we masked out parts of the spectrum contaminated by
telluric lines.  Although the star appeared as a fast rotator with a \vsini\ value of 
about 17~\kms , these first radial velocity measurements allowed us 
to confirm the planetary nature of the transiting body.

\begin{table}[h]
\begin{center}{
\caption{Radial velocity measurements for \corotp.}
\label {VRadMeas} 
\begin{tabular}{lcccl}
\hline
\hline
HJD    & exp time & \vrad\  & $\sigma$\vrad  & Spectrograph \\
          &  (sec)     & \kms   &  \kms                &    \\
\hline
2454309.5696  & 3600 & $-$58.262 & 0.046 & \sophie\ \\
2454316.4672  & 2700 & $-$54.269 & 0.060 & \sophie\ \\
2454317.4443  & 3000 & $-$57.122 & 0.060 & \sophie\ \\
2454318.3674  & 3000 & $-$58.371 & 0.078 & \sophie\ \\
2454336.4002  & 2700 & $-$55.938 & 0.047 & \sophie\ \\
2454590.6002  & 3600 & $-$58.386 & 0.051 & \sophie\ \\
2454643.5273  & 2700 & $-$54.386 & 0.074 & \sophie \\ 
2454331.3590  & 3600 & $-$57.130 & 0.250 & \tls \\ 
2454332.3600  & 3600 & $-$54.460 & 0.290 & \tls \\
2454336.9060  & 3600 & $-$53.650 & 0.240 & \tls \\
2454341.6815  & 1800 & $-$53.978 & 0.052 & \harps\ \\
2454342.6366  & 1800 & $-$55.867 & 0.049 & \harps\ \\
2454343.6556  & 1800 & $-$58.331 & 0.054 & \harps\ \\
2454344.6816  & 1800 & $-$56.683 & 0.057 & \harps\ \\
2454320.7188  & 3600 & $-$54.628 & 0.200 & \coralie\ \\
2454322.7074  & 3600 & $-$58.523 & 0.198 & \coralie\ \\
2454330.6931  & 3600 & $-$58.457 & 0.141 & \coralie\ \\
2454372.4932  & 3600 & $-$56.263 & 0.172 & \coralie\ \\
2454372.5133  & 3600 & $-$56.597 & 0.186 & \coralie\ \\
2454372.5281  & 3600 & $-$56.565 & 0.218 & \coralie\ \\
2454372.5429  & 3600 & $-$56.644 & 0.170 & \coralie\ \\
2454372.5629  & 3600 & $-$56.535 & 0.189 & \coralie\ \\
2454372.5829  & 3600 & $-$57.039 & 0.210 & \coralie\ \\
2454372.6032  & 3600 & $-$57.154 & 0.221 & \coralie\ \\
2454372.6231  & 3600 & $-$57.265 & 0.259 & \coralie\ \\
2454372.6433  & 3600 & $-$56.951 & 0.214 & \coralie\ \\
2454379.5247  & 3600 & $-$54.772 & 0.178 & \coralie\ \\
2454380.5267  & 3600 & $-$55.498 & 0.192 & \coralie\ \\
2454387.5000  & 3600 & $-$56.494 & 0.217 & \coralie\ \\
2454388.5207  & 3600 & $-$54.844 & 0.226 & \coralie\ \\
2454394.5333  & 3600 & $-$58.696 & 0.211 & \coralie\ \\
2454400.5461  & 3600 & $-$55.556 & 0.232 & \coralie\ \\
2454607.8861 & 3600  & $-$57.786 &  0.163  & SANDIFORD \\
2454609.8528 & 3600  & $-$53.187 &  0.100 &  SANDIFORD\\
2454613.8684 & 3600   & $-$53.308  &  0.157 &  SANDIFORD\\
\hline
\end{tabular}}
\end{center}
\end{table}

Four complementary measurements were carried out with the \harps\ spectrograph 
on the 3.6m telescope (La Silla, Chile) in order to perform the line-bisector analysis to look 
for asymmetries in the spectral line profiles, as could be caused by contamination from 
an unresolved eclipsing binary \citep{2001A&A...379..279Q}. 
Four exposures of 30~min each were recorded over four consecutive nights covering one orbital period. 
Eighteen additional measurements were carried out with the \coralie\ spectrograph 
on the Euler 1.2-m telescope. \coralie\ has recently been upgraded, as described in \cite{2008ApJ...675L.113W}, in order to follow-up \corot\ targets up to 14th magnitude. 

Further spectra were collected at McDonald Observatory (Texas, USA)
during three nights on May 2008 in order to investigate a possible 
drift in the radial velocity. 
The Sandiford echelle spectrograph mounted at the Cassegrain 
focus of the 2.1m Otto Struve telescope was used in conjunction with 
a $1^{\arcsec}$ slit. The configuration yielded 
a spectral coverage of 5070-6050~{\AA} at a 
resolving power $R=60 000$. Two consecutive spectra of 30 minutes 
were obtained during each night and with 
calibration spectra acquired
before and after each stellar observation.
Spectral order extraction followed 
standard procedures under the IRAF environment. HR\,5777 was used
as the standard star for the radial velocity 
cross-correlation.  

  \begin{figure}
   \centering
    \includegraphics[height=6cm,width=9cm]{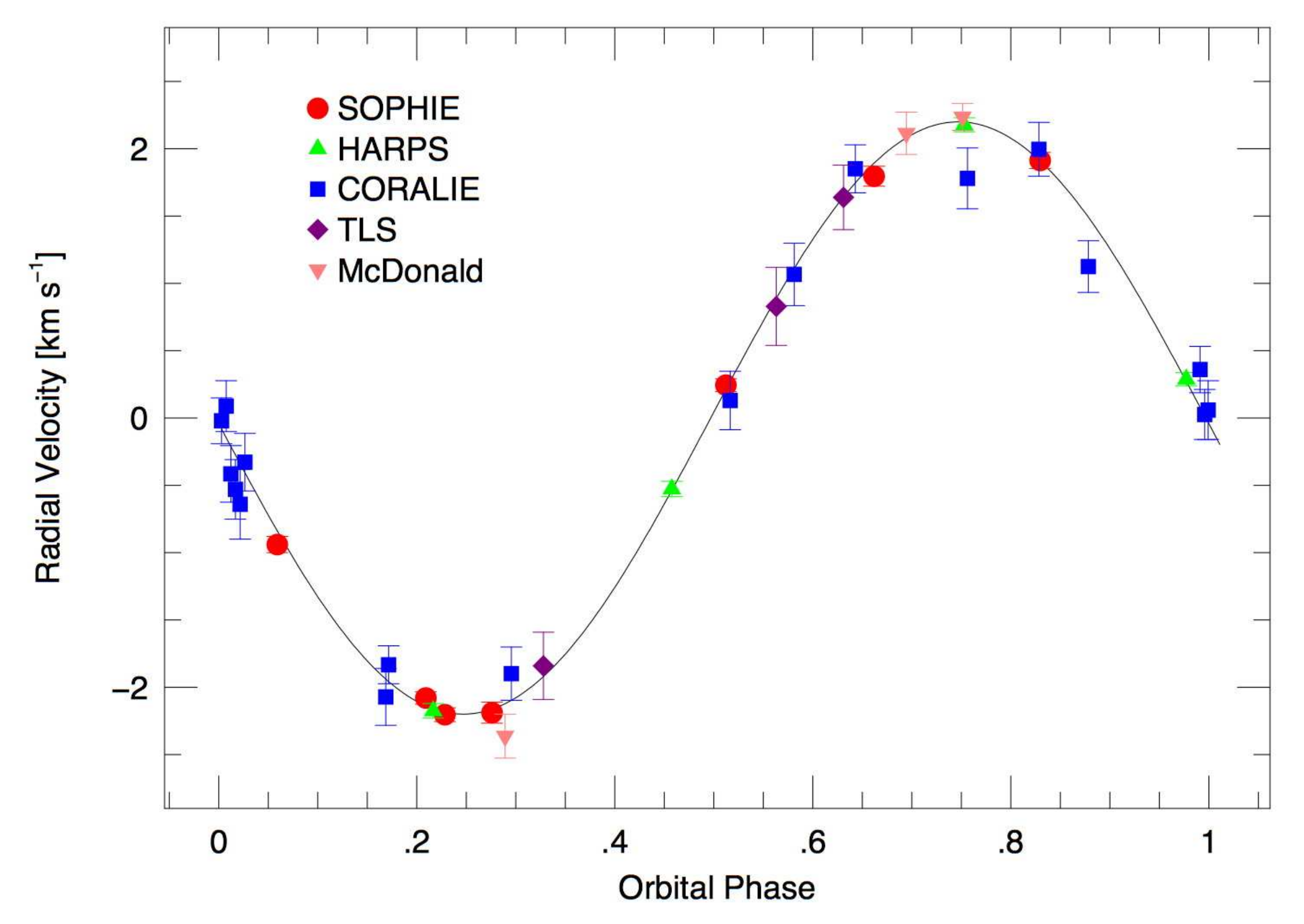}
   \caption{The phase-folded radial velocity measurements from the different spectrographs used for the follow-up campaign,
with the best fit solution over-plotted (solid line).  }              
\label{RadVel}
    \end{figure}
\begin{figure}[]
\begin{center} 
\includegraphics[height=4cm, width=9cm]{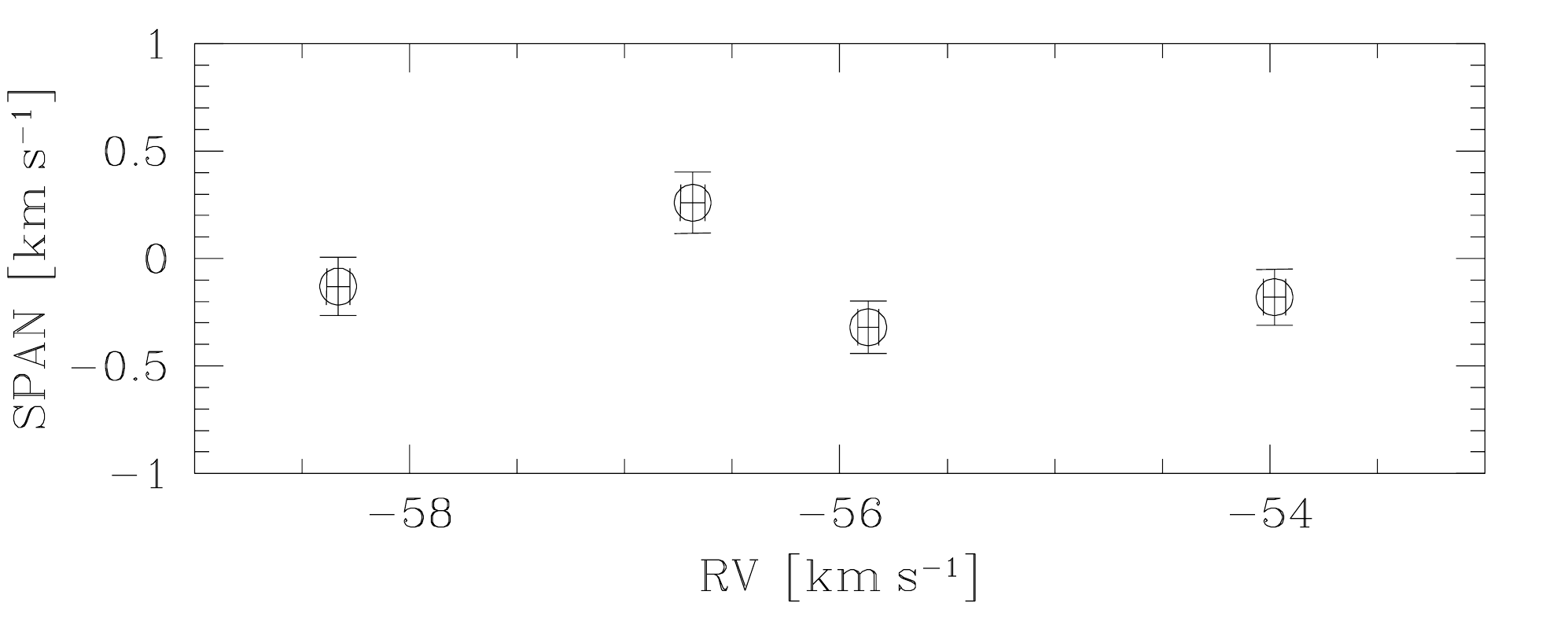}
\end{center}
\caption{\label{Bisector}The bisector span, giving the differences in radial velocities between top and bottom of a bisector, as a function of radial velocity, measured with \harps.}
\end{figure}
  
The complete radial velocity measurements obtained with these five  spectrographs are presented in Table~\ref{VRadMeas}. The five  sets of relative radial velocities were simultaneously fitted with a Keplerian model, with the epoch of the transit being fixed at the \corot\ value and with an adjusted offset between the different instruments. 
No significant eccentricity was found and we decided to set it to zero. 
The best fit parameters yields K = 2.194 \kms\ $\pm$ 0.027.
Fig~\ref{RadVel} shows all the  radial velocity measurements after
subtracting the individual offsets and phase folded to the orbital period.

A line-bisector analysis is routinely done for each \corot\ candidate in order to identify the presence of any spatially unresolved stellar companion which could be the source of the radial velocity variation. This analysis performed on the 4 \harps\ measurements of our target does not show any significant variation (see Figure~\ref{Bisector}) thus
excluding a blend scenario due to an unresolved eclipsing binary. Several \coralie\ measurements were made during the transit but the Rossiter-McLaughlin effect was not detected. 
Indeed, considering the planet-star radius ratio of 0.066 and the rotational velocity of \vsini\ of 17~\kms, the expected semi-amplitude of the radial velocity anomaly for a central and spin-aligned transit is only 55~\ms. This is well below the \coralie\ radial velocity uncertainties of $\sim$ 200 \ms ; a further such measurement with \harps\ is going to be carried out. 

\begin{figure}[]
\begin{center} 
\includegraphics[height=5.5cm, width=9cm]{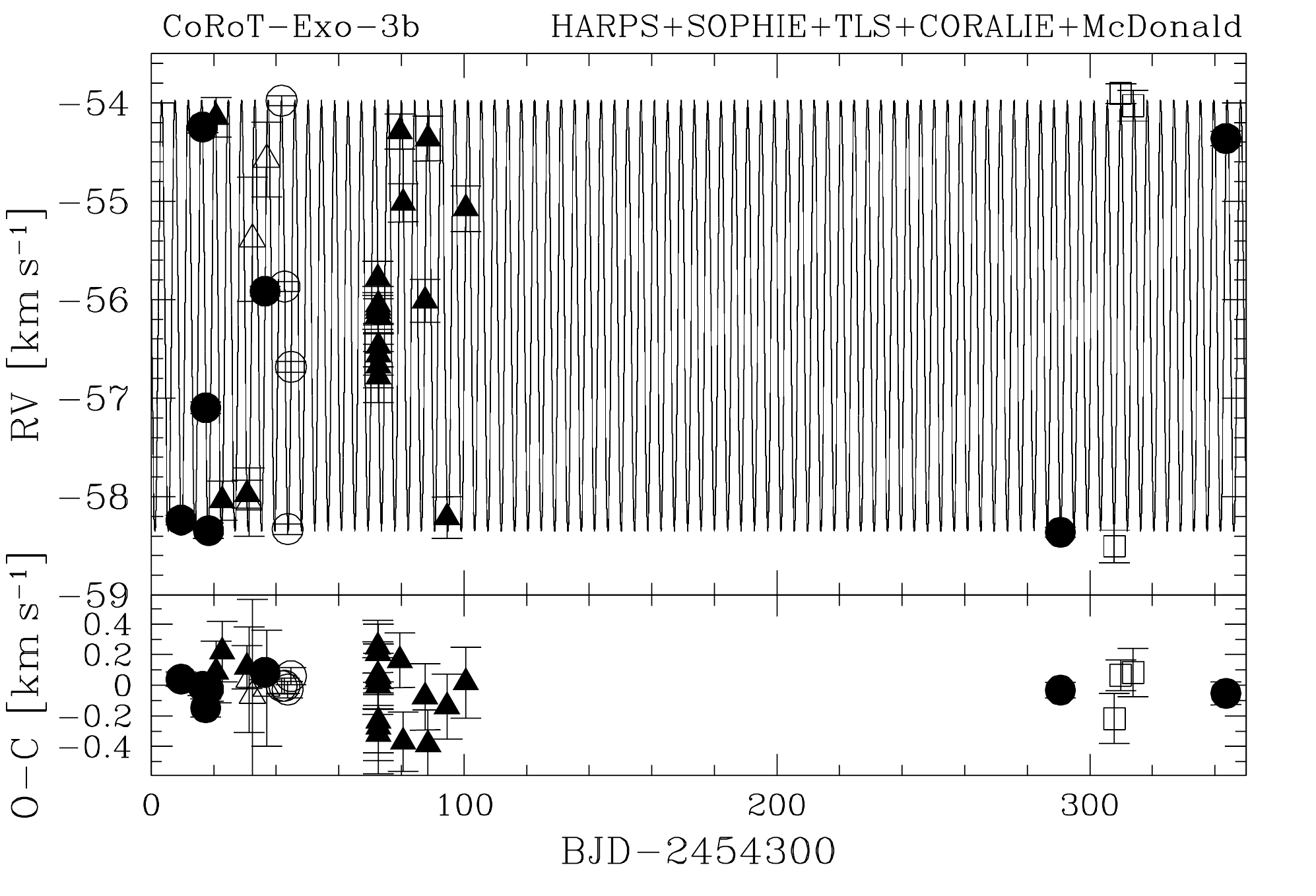}
\end{center}
\caption{\label{VradTot}The complete set of radial velocity measurements over a  one year period: \sophie\ (black circle), \tls\ (open triangle), \harps\ (open circle), \coralie\ (black triangle) and  {\sl SANDIFORD} (open square) spectrographs. 
}
\end{figure}

In total, the radial velocity measurements cover a duration of more than 11 months, five additional measurement recently
made on Spring 2008 (Fig~\ref{VradTot}). No significant drift was detected over this period, excluding the presence of an additional massive companion ($>$ 2 M$_{Jup}$) with a period less than 11 months.  

Considering the projected rotation velocity \vsini\ of 17~$\pm$~1 \kms and assuming that the rotation axis of the star is perpendicular to the line-of-sight, the star is rotating in 4.6~$\pm$~0.4 days. 
Given the relatively large mass and short period of the companion, one might expect it to have synchronized the rotation of its host star to its orbital period via tidal interaction \citep{2004ApJ...610..464D,2008ApJ...681.1631J}. The observed rotational velocity of the star is indeed compatible with this hypothesis and the phenomenon has recently been reported for two other massive planets orbiting an F-star in circular orbit: $\tau$ Boo \citep{2008MNRAS.385.1179D} and \corotfour\ \citep{2008A&A...488L..43A}. We attempted to measure the photometric rotation period of the star from the out-of-transit light curve, using a discrete autocorrelation function (ACF) method 
as was successfully done for \corotfour\ \citep{2008A&A...488L..43A}. However, no significant signal was  detected, even when computing the ACF using only the second half of the light curve, where the photometric variability is slightly more pronounced. It appears that the star is not sufficiently active to sustain coherent active regions producing detectable rotational modulation over more than one period. 

\subsection{High resolution spectroscopy} 
\label{SpectroParam}
The \sophie\ and \harps\ spectra
were of too poor a quality to allow a proper spectral analysis. 
Consequently, the star was observed with the \uves\ spectrograph at the end of October 2007. Two exposures of 2380~sec each were recorded, using the a spectrometer slit of 0.7~\arcsec\ which yielded a resolving power of $\simeq$ 65~000. The co-added spectra gives a \sn\ ratio greater than 140 per resolution element over the entire spectral range.

\begin{table}[h]
\begin{center}{
\caption{\label{StarParam} 
\corots\  parameters derived from radial velocity and spectroscopic analyses.}
\begin{tabular}{lll }\hline
\hline
\vrad\ (\kms)  & $-$56.162  & $\pm$ 0.016 \\
$v_{\rm rot} \sin i$ (\kms) &  17.0      &  $\pm$ 1.0\\  
\teff\ & 6740K & $\pm$ 140 \\
\logg\ & 4.22~$^{\mathrm{(a)}}$ & $\pm$ 0.07 \\
\logg\ & 4.25~$^{\mathrm{(b)}}$ & $\pm$ 0.07 \\
\met\ &  $-$0.02 & $\pm$ 0.06 \\
Spectral Type & F3 V\\
$M_\star$ & 1.37 &  $\pm$ 0.09 \\
$R_\star$ &  1.56  & $\pm$ 0.09 \\
Age  & 1.6 - 2.8 Gyr & \\
Distance &  680 pc &  $\pm$ 160 \\
\hline
\end{tabular}}
\begin{footnotesize}
\begin{list}{}{}
\item (a) Determined from the spectroscopic analysis
\item (b) Determined using evolutionary models and the light-curve \mr\ parameter.
\end{list}
\end{footnotesize}
\end{center}
\end{table}

The spectral analysis was performed using different methods and 
by different \corot\ teams in an independent way. 
Some of the methods consist of spectral synthesis modeling, using a library 
of synthetic spectra, as presented in \cite{2008A&A...482L..17B}. In particular, we used 
the Spectroscopy Made Easy (SME 2.1) package \citep{1996A&AS..118..595V,2005yCat..21590141V}.
Other methods are based on equivalent width measurements of isolated lines, while the semi-automatic software 
package VWA \citep{2002A&A...389..345B,bruntt08} performs iterative fitting of calculated synthetic spectra. These methods required a careful normalization of the spectra. 
This was done order per order by fitting a spline function to the continuum windows  identified in a synthetic spectrum and calculated with parameters close to those of the target star. 
We made sure that the shape and depth of lines in adjacent spectral orders were in agreement. 
In the final spectrum the orders were merged and the overlapping parts were weighted by the \sn. 
We found that the estimated values of the atmospheric parameters: effective temperature (\teff),  
surface gravity (\logg) and metallicity (\met), obtained by the different methods agreed within 
the estimated errors. The adopted values are listed in Table~\ref{StarParam}.
 
As part of the analysis we adjusted the macroturbulence to fit the wings 
of the spectral lines and found $v_{\rm macro}$ = 4.0 $\pm$ 0.6~\kms. 
The rotational broadening was found to be $17\pm1$\,\kms.

The detailed abundance results we report here are based on the VWA method.
We used 192 mostly non-blended lines and the abundances were calculated 
relative to the solar spectrum from \cite{hinkle00}, 
following the approach by \cite{bruntt08}.
Using the fundamental parameters listed in Table~\ref{StarParam} 
for the atmospheric model, we determined the abundances 
of 17 individual elements as listed in Table~\ref{Tabvwa0523}. 
 
\begin{table}
 \centering
 \caption{\label{Tabvwa0523}Abundances of 17 elements in \corots.}
\begin{tabular}{l l c}
\hline\hline
Element & $\log{N / N_{\rm tot}}$ & number \\
             &                                      & of lines \\
\hline
  {C  \sc   i}   &  $ -0.40      $  &   1  \\
  {Na \sc   i}   &  $ -0.14      $  &   2  \\
  {Mg \sc   i}   &  $ -0.08      $  &   1  \\
  {Si \sc   i}   &  $ -0.03(0.07)$  &   3  \\
  {Ca \sc   i}   &  $  +0.03(0.07)$  &   5  \\
  {Sc \sc  ii}   &  $ -0.11(0.08)$  &   4  \\
  {Ti \sc   i}   &  $ -0.01(0.08)$  &   8  \\
  {Ti \sc  ii}   &  $  +0.02(0.06)$  &   5  \\
  {V  \sc   i}   &  $ -0.36      $  &   2  \\
  {Cr \sc   i}   &  $ -0.09(0.11)$  &   5  \\
  {Cr \sc  ii}   &  $ -0.06(0.07)$  &   5  \\
  {Mn \sc   i}   &  $ -0.26(0.08)$  &   7  \\
  {Fe \sc   i}~$^{\mathrm{(a)}}$ &  $ +0.03(0.06) $  &  92  \\
  {Fe \sc  ii}   &  $  +0.02(0.06)$  &  13  \\
  {Co \sc   i}   &  $  +0.04      $  &   2  \\
  {Ni \sc   i}   &  $ -0.06(0.06)$  &  29  \\
  {Cu \sc   i}   &  $ -0.66      $  &   1  \\
  {Zn \sc   i}   &  $ -0.18      $  &   1  \\
  {Y  \sc  ii}   &  $ -0.07(0.20)$  &   4  \\
  {Zr \sc  ii}   &  $ -0.15      $  &   2  \\
\hline
\end{tabular}
\begin{footnotesize}
\begin{list}{}{}
\item (a) corrected for  non-LTE effects
\end{list}
\end{footnotesize}
\end{table}

\begin{figure}[]
\begin{center} 
\includegraphics[height=4cm, width=9cm]{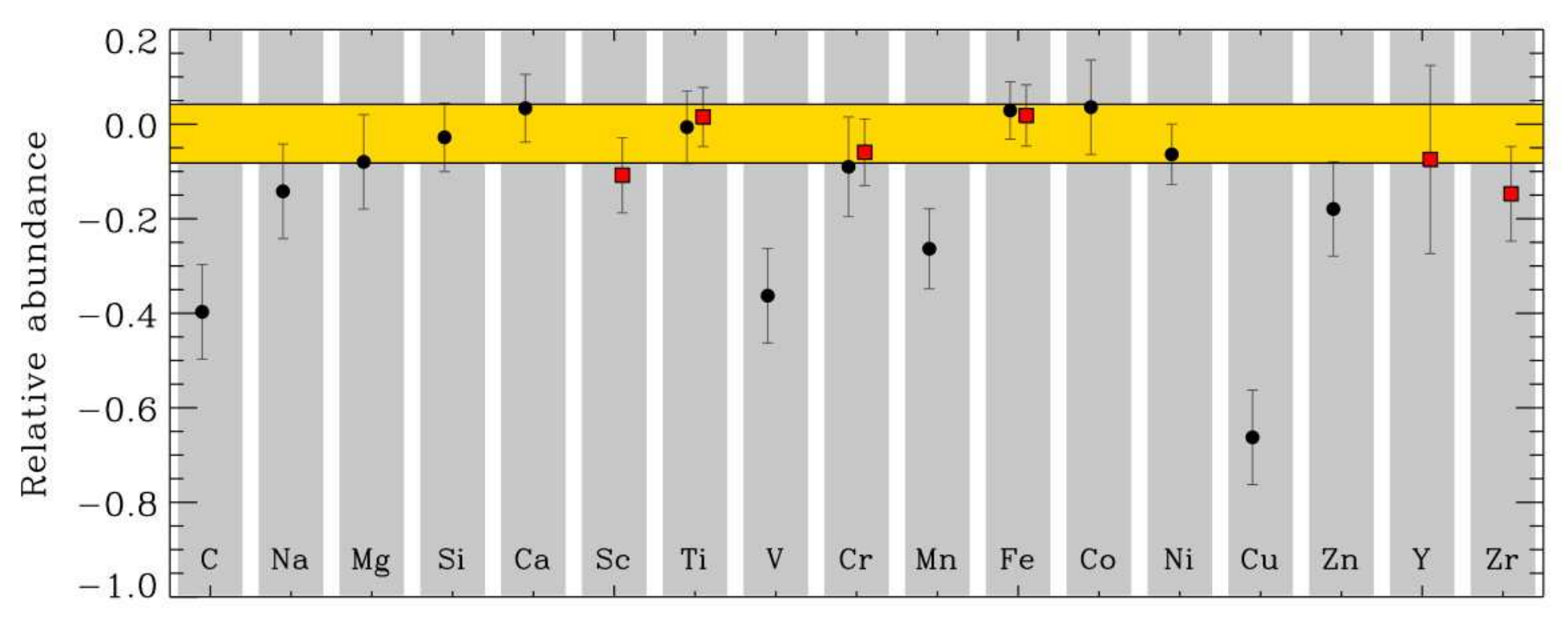}
\end{center}
\caption{\label{vwa0523}Abundances of the elements measured in \corots. Black circles are for neutral elements and box symbols are used for ionized ones.
The metallicity and the 1-$\sigma$ uncertainty range are indicated by
the horizontal yellow bar.}
\end{figure}

The uncertainty on the abundances includes a contribution of $0.06$ dex
due to the uncertainty on the fundamental parameters.
The overall metallicity is found as the mean abundance of the elements
with at least 5 lines, that is Ca, Ti, Cr, Fe and Ni, giving [M/H]\,$=-0.02\pm0.06$  (Fig.~\ref{vwa0523}). 
We did not include Mn since it has a significantly lower
abundance than the other metals. It is worth noticing that 
we found no evidence for the star being chemically peculiar.
Our target star is slightly cooler than this ($T_{\rm eff}=6740$\,K) and departure from LTE is expected. 
\cite{holm96} investigated the NLTE effects for Fe
in A-type stars with $T_{\rm eff} > 7000$\,K. 
We extrapolated from her Fig.~4 for solar metallicity
to find a first-order correction
[\ion{Fe}{i}/H]$_{\rm NLTE}$ = [\ion{Fe}{i}/H]$_{\rm LTE}+0.05$. This value has been added to \ion{Fe}{i} in Table~\ref{Tabvwa0523}.
The abundances found from \ion{Fe}{ii} lines are essentially unaffected.
To make the atmosphere model produce the same result for neutral and 
ionized Fe lines, $\log g$ was increased by $0.10$~dex.

\begin{figure}[]
\begin{center} 
\includegraphics[height=5cm, width=7cm]{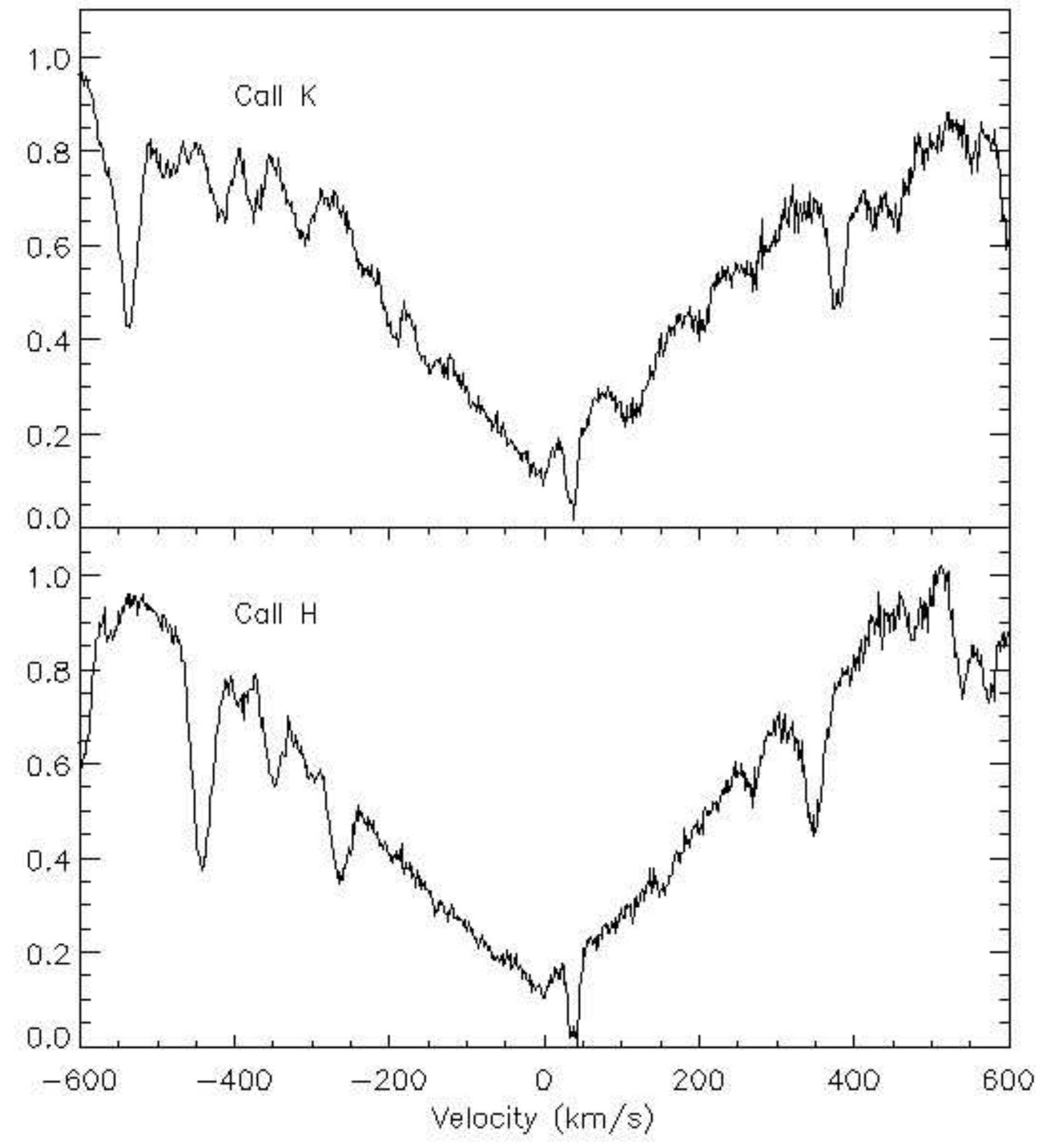}
\end{center}
\caption{The \ion{Ca}{ii} H \& K lines observed in the \uves\ spectra of \corots. No emission feature is seen is the core of the stellar lines. The velocity scale is in the rest frame of the star.}
\label{Caii523}
\end{figure}

Inspection of the spectra reveals no emission in the \ion{Ca}{ii} H and K lines (Fig.~\ref{Caii523}) or other photospheric lines.
This is in good agreement with the lack of photometric variation in the light curve, as well as with no strong jitter in the radial velocity measurements.

\begin{figure}[]
\includegraphics[height=6cm, width=8.5cm]{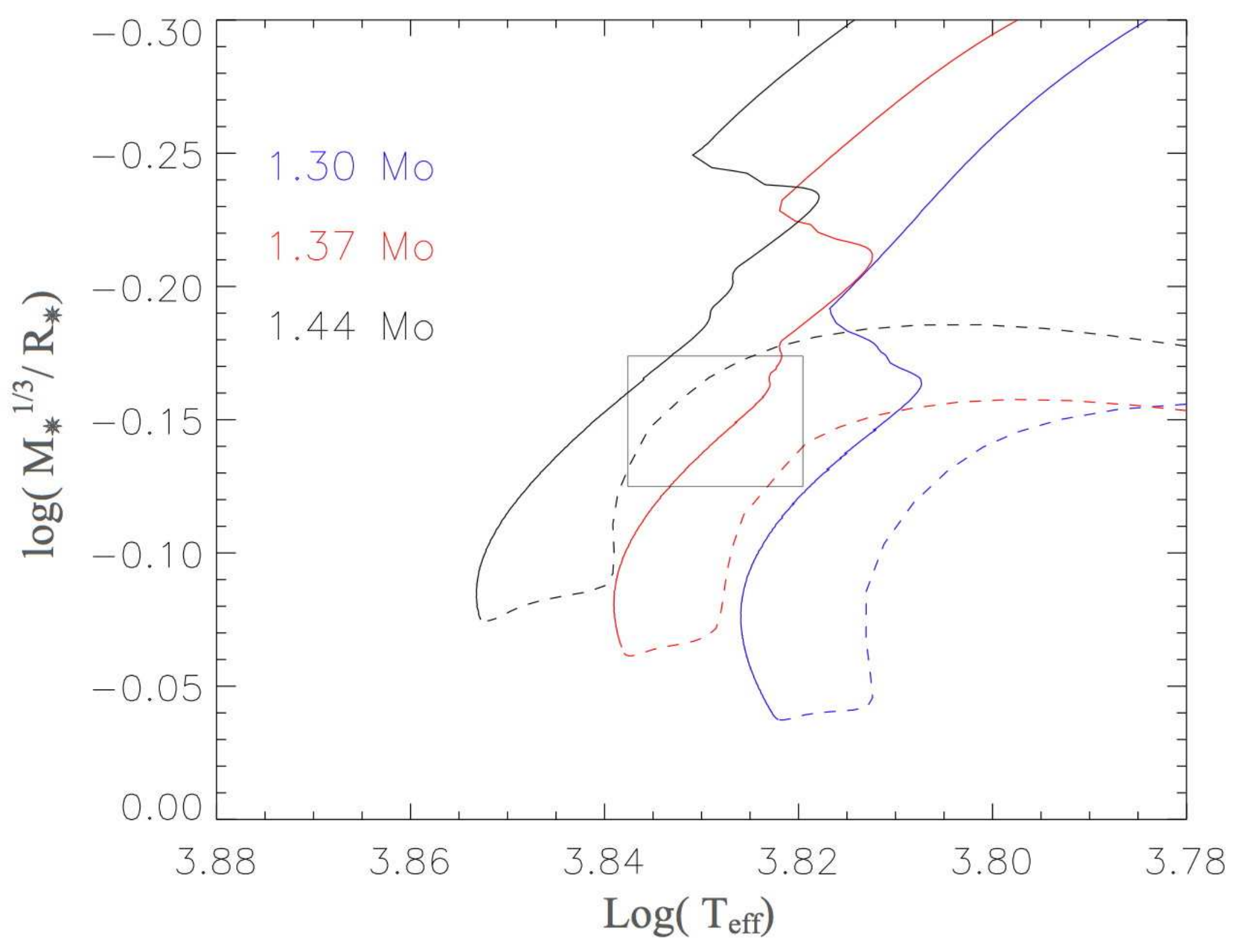}
\caption{\label{HRplot} Stellar evolutionary tracks from the STAREVOL models} for masses in the range 1.30 to 1.44 $M_\odot$ for the measured 
metallicity [M/H]\,$=-0.02$, shown with the parameters derived from the observations: \teff\ and \mr.
\end{figure}

To determine the mass and radius of the parent star we
used the same methodology as for the two first \corot\ planets \citep{2008A&A...482L..17B,2008A&A...482L..25B},
i.e.\ we used \teff\ and [M/H] from the spectroscopic analysis and \mr\ from the light curve analysis which provides a better estimate of the fundamental parameters, thanks to the good quality of the \corot\ light curve. From a comparison with evolutionary models as shown in Fig.~\ref{HRplot}, we can constrain the fundamental parameters of the parent star. In this study, we mainly relied on {\sl STAREVOL} (\citealt{2006A&A...448..717S}; Palacios, {\sl private communication}) stellar evolution models to derive the stars precise parameters. We also compared these with  the results obtained using  {\sl CESAM} \citep{2007Ap&SS.tmp..460M} and we found that both stellar evolution model tracks were in agreement. The details of the comparison between the different models will be presented in a forthcoming paper. 
We find the stellar mass to be $M_\star = 1.37 \pm 0.09\,M_{\sun}$ and the stellar  radius $R_\star = 1.56 \pm 0.09$ R$_{\sun}$, with an age in the range $1.6$--$2.8$~Gyr.
This infers a surface gravity of $\log g = 4.24\pm0.07$, 
which is in good agreement with the spectroscopic result of 
 $\log g = 4.22\pm0.07$,
and implies that the correction due to NLTE effects is indeed relevant.

In a final step, we calculated the distance to the star. We used the physical parameters of the star we derived and its colors to estimate the reddening. Using the extinction law from \cite{1985ApJ...288..618R}, we found the absorption $A_V = 0.52~\pm 0.5$~mag, yielding a distance of $680~\pm 160$~pc. We checked that the value of the extinction we derived is consistent with strong saturated \ion{Na}{i} (D1) and (D2) interstellar lines as well as reddening maps of \cite{1998ApJ...500..525S} which give a maximum absorption towards our target of $1~mag$.  

We also investigated the possibility of a physical association between \corots\ and the nearby brightest companion at 5.6~\arcsec. Given our estimated distance to
the star, the range of possible extinction on the line of sight, and the apparent visual magnitude of the contaminant ($V = 16.46 \pm 0.07$) we derived an absolute visual magnitude $M_V = 6.8 \pm  0.5$, consistent with a K-type star. We compared the colors of the contaminant we calculated from our ground-based observations and {\mass} photometry (Table~\ref{StarID}) with those predicted for a star of this spectral type. We found that within the precision of the different parameters, a physical association could not be excluded. In that case, their separation would be about 3800~AU and the orbital period of the companion would be $\sim$ 235 000 years. On the other hand, according to the \ucac\ catalogue \citep{2004AJ....127.3043Z}, \corots\ displays a proper motion while none is detected for the companion. 
This non detection would hence rather favor a background star.   
Given our current knowledge we can not draw a firm conclusion about the
 possibility of  binarity for \corots. More complementary observations
are required.

\section{CoRoT-exo-3b parameters and discussion}
\subsection{Nature of CoRoT-exo-3b}
\begin{figure}[]
\begin{center} 
\includegraphics[height=6.5cm, width=8.5cm]{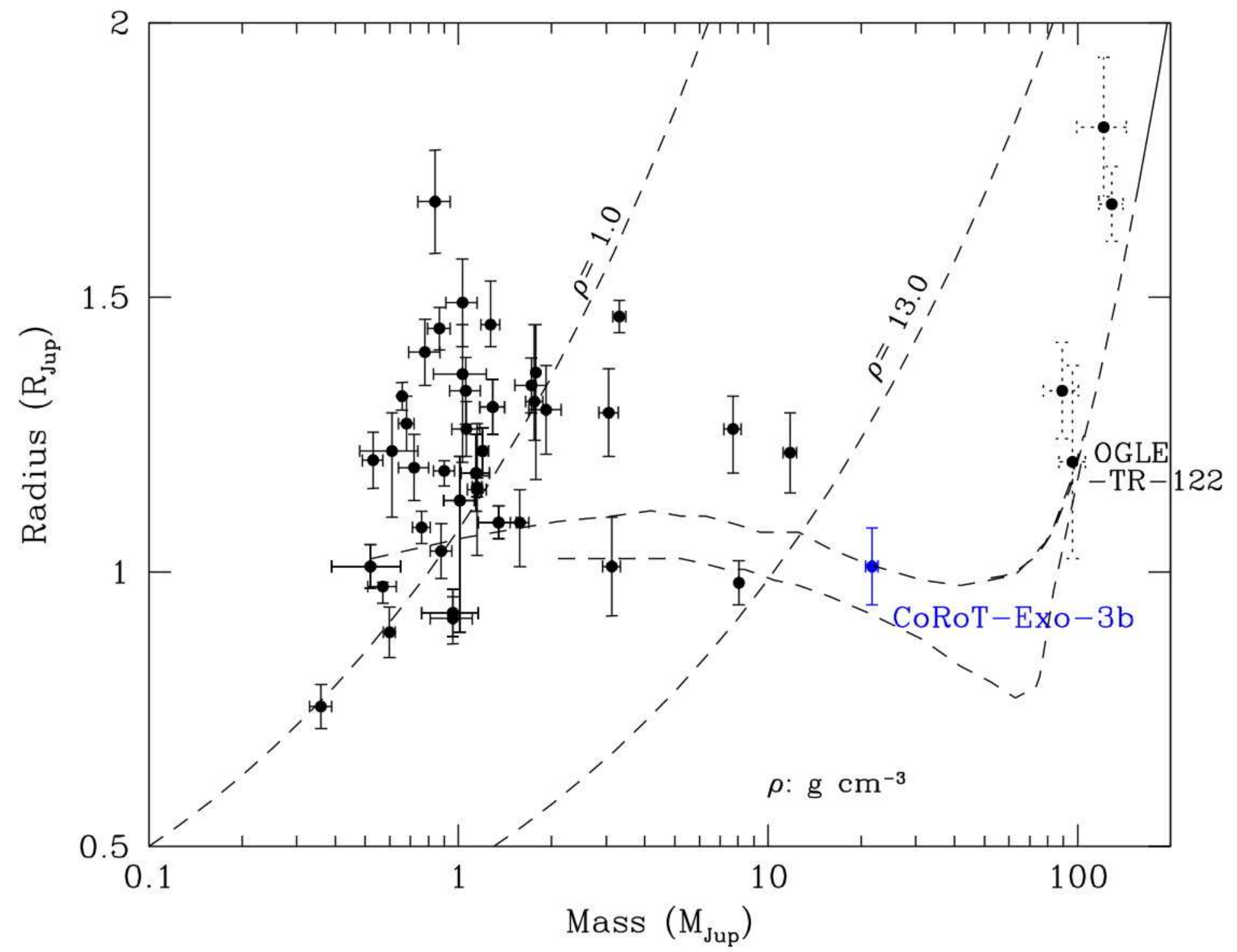}
\end{center}
\caption{\label{MassRadiusHR} Mass-radius diagram for all transiting planets and low-mass M stars with theoretical isochrones at 10 and 1 Gyr from \cite{2003A&A...402..701B} overplotted.}
\end{figure}

Using the stellar properties determined in the previous section and the characteristics of the transiting body as derived from the transit and the radial velocity fits, we derive a mass of the companion of  $M_p$ = 21.66 $\pm 1.0$ {\MJ}, a radius $R_p$ = 1.01 $\pm 0.07$ {\RJ},  an inferred density of 
$\rho$ = 26.4~$\pm$~5.6~\gc, and a surface gravity of \logg = 4.72~$\pm$~0.07 (Table~\ref{BD_ID}). With such properties, \corotp\ clearly distinguishes itself 
from the regular close-in extrasolar planet population. In a mass-radius diagram, the position of \corotp\ is well inside the gap in mass between planetary and low-mass star companions (Fig~\ref{MassRadiusHR}). 
\begin{table}[h]
\caption{\label{BD_ID} 
\corotp\ parameters.}
\begin{center}{
\begin{tabular}{lr }
\hline
\hline
Mass (\MJ)& 21.66 $\pm$ 1.0 \\
Radius   (\RJ)  &  1.01 $\pm$ 0.07 \\ 
density  (\gc)  &   26.4  $\pm $5.6   \\
\logg\       &   4.72 $\pm$ 0.07 \\
\hline
\end{tabular}}
\end{center}
\end{table}

Traditionally, a planet has been defined as an object lighter than 13~M$_{Jup}$, as such objects are supposed not to have 
an internal nuclear source of energy
 (Deuterium burning). From this point of view, \corotp\ is definitely a brown-dwarf. Indeed, in this low mass range, models predict an almost constant Jupiter-size radius \citep{2003A&A...402..701B}. 
 As illustrated in  Fig~\ref{MassRadiusHR}, \corotp\ parameters are in good agreement with the expected mass-radius relationship on the low-mass tail of these substellar objects. 

 Another definition makes use of the formation scenario: a planet is formed by core accretion of dust/ices in a  protoplanetary disk, while a brown-dwarf is formed by collapse of a dense molecular gas cloud.  In that case, the separation between the brown-dwarf and planet population is blurrier since a planet, 
starting with a  solid core, can end up with a gaseous envelope as massive as
a few tens of Jupiter masses. Recent improvements of the core-accretion models predict the formation of a wide variety of giant planets \citep{2005A&A...434..343A} with masses up to 10~\MJ, depending on the initial conditions. Some authors \citep{Mordasini07} even suggested that the mass of these ``superplanets'' could be as high as 25~\MJ. An alternative hypothesis for 
the origin of such massive planetary bodies could be collisions between several massive planets, as recently proposed by \cite{2008A&A...482..315B}. 

\corotp\ could be considered either as a member of a new population of very massive planets, or ``superplanets", or a representative of the low-mass part of the brown-dwarf family. 

\subsection{Is \corotp\ exceptional?}
The discovery of a 21~\MJ\ object in a short period
orbit is unexpected given the paucity of objects in the  so-called 
``brown-dwarf desert'' \citep[e.g.][]{2000A&A...355..581H,2006ApJ...640.1051G}. However, Doppler surveys have found companions to HD~41004B \citep{2004A&A...426..695Z} and HD~162020 \citep{2002A&A...390..267U} that bear some 
similarities with  \corotp\  with  a minimum  masses between 10 and 20 M$_{Jup}$ and  short orbital periods. Nevertheless, it is difficult to assess the statistical significance of the lack of short period companions with masses between 10 and 20 M$_{Jup}$ detected from the radial velocity surveys. For the very large (over 20\, 000 stars) Doppler mid-accuracy survey made by {\sl CORAVEL} \citep{2004A&A...418..989N}, the lack of sensitivity may have prevented these surveys from detecting such companions orbiting fast rotating stars.
Accurate Doppler planet surveys, based on a much larger and complete sample, like the one carried out with \coralie\ \citep{2000A&A...356..590U} would have the capability to detect a companion like \corotp. However, in that case, the observing strategy is focused on slow rotators. The significant rotational line broadening  would have affected the search strategy, either by removing these objects after one measurement or by setting them aside with a very low observation ranking priority. This selection bias towards slow rotators may now become remedied, thanks to recent studies aiming at enlarging the space of parameters for host stars \citep[e.g.][]{2005A&A...443..337G}.

\begin{figure}[]
\includegraphics[height=6cm, width=8.5cm]{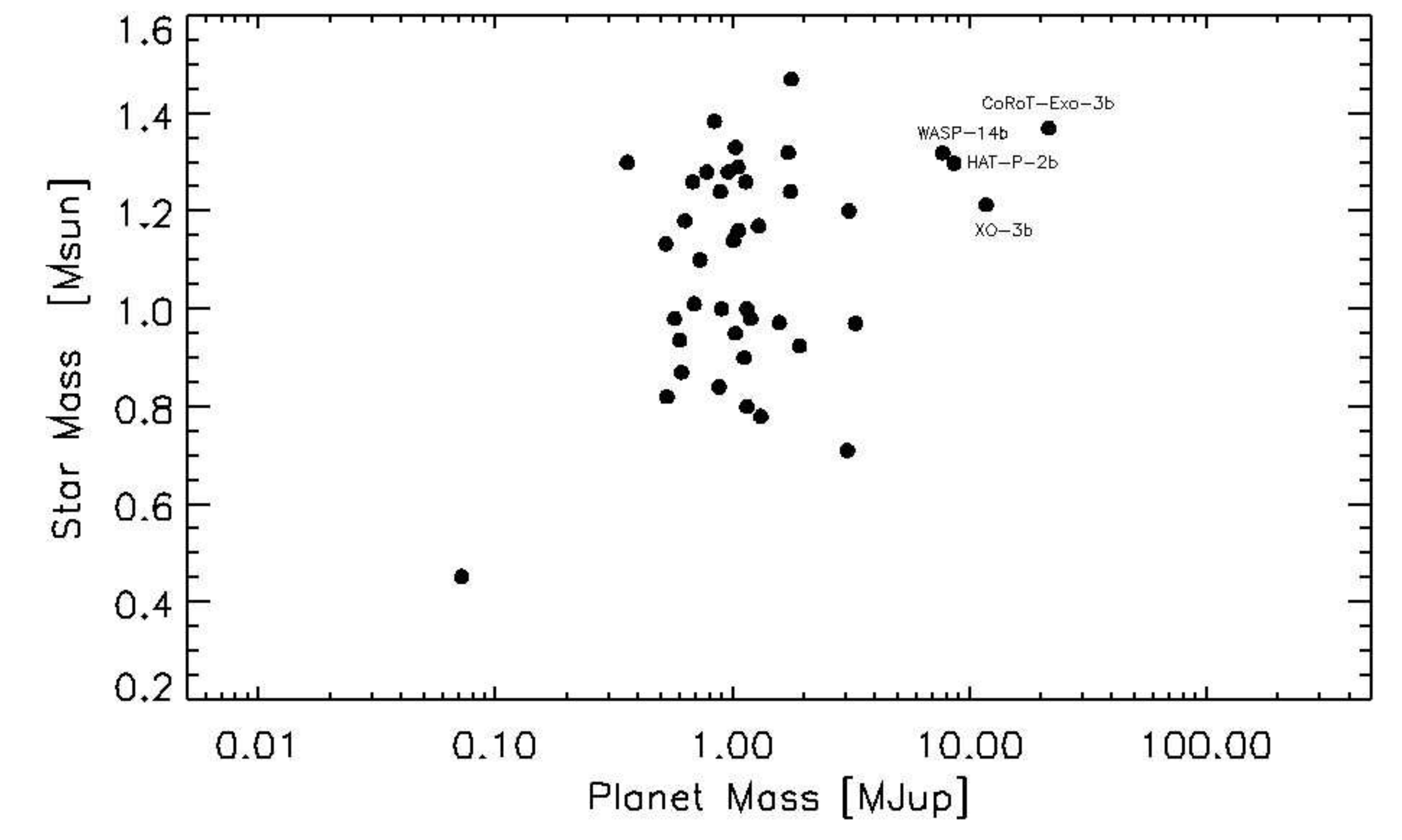}
\caption{Mass of the known transiting extrasolar planets with orbital period less than 10 days, as a function of the mass of their parent star}
\label{StarPlanet}
\includegraphics[height=5cm, width=8.5cm]{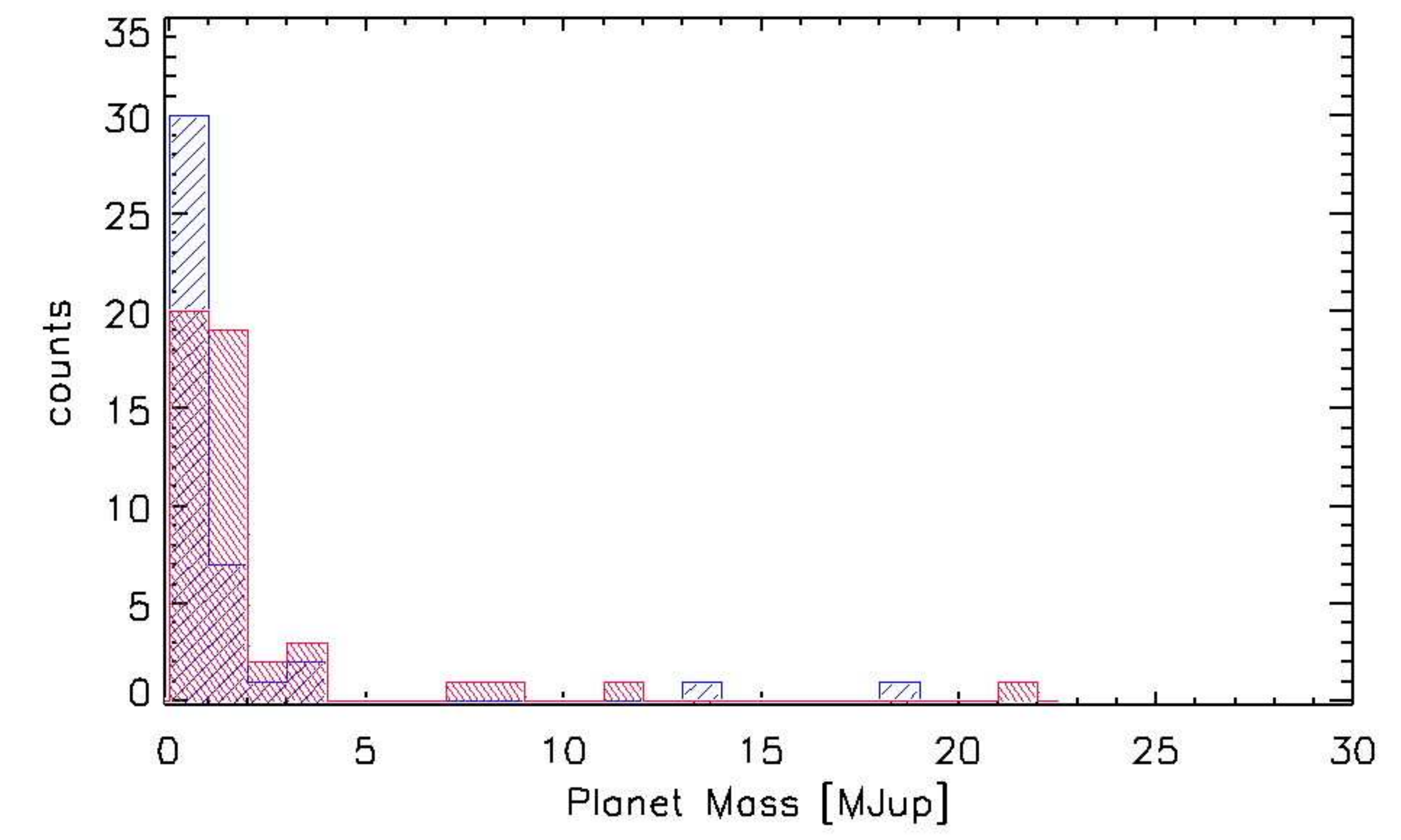}
\caption{\label{Masshisto} Mass distribution of close-in planets. Blue: planets with orbital period less than 30 days, detected by radial velocity and with a minimum mass estimate only ; red : transiting planets}
\end{figure}

\subsection{Do massive companions require massive stars?}

Among the 46 well characterized transiting planets, 19 are orbiting a star with a mass $\ge$ 1.1~M$_{\sun}$, as illustrated in Fig~\ref{StarPlanet}. 
It is worth noticing that the three most massive planets so far discovered are orbiting F-type stars: HAT-P-2b \citep{2007ApJ...671L.173B} with mass of 8.6 M$_{Jup}$ orbits a F8 star, WASP-14b \citep{2008arXiv0806.1478J} and XO-3b \citep{2008ApJ...677..657J}  with masses of 7.77~\MJ\ and 11.79~\MJ\ respectively, both
orbiting  F5-type stars. Although the sample is still very limited, it suggests that the close-orbiting companions to host stars with mass above 1.1 M$_{\sun}$ could be more massive than companions to lower mass hosts. In that case, we could not exclude that the brown-dwarf desert is not that dry around F-type stars.

Alternatively, the discovery of companions in short periodic orbits with masses between 5 and 20~M$_{Jup}$ could be suggestive of a different formation mechanism. The mass distribution of planet companions in short period orbits found by radial velocity surveys on solar type stars becomes sparsely populated for masses beyond 3~M$_{Jup}$ (Fig.~\ref{Masshisto}). Below 3~M$_{Jup}$ the planet population is steeply rising as a function of decreasing planet mass. In a mass distribution diagram these massive planets look clearly off the short tail of the planet distribution. Different formation mechanism  for massive planets on short orbits have already been pointed out by \cite{2002A&A...390..267U}, who noticed that they seem to be always found in binary systems. In the case of {\corots}, more investigations may be necessary to establish unambiguously whether the star belongs to a binary system or not. If one compares the bulk of Doppler planets with the metallicity distribution of HAT-P-2, WASP-14, XO-3  and  \corots, it is interesting to point out that it is not skewed towards metal rich stars, as one would expect for such planets \citep{2007ARA&A..45..397U}, suggesting again 
 possibly a different formation mechanism. Therefore one might conclude that \corotp\ could be different from the bulk of extrasolar planets found by Doppler surveys but not from the few companions with masses above 3~M$_{Jup}$. 
On the other hand, we note that HAT-P2b, WASP-14b and XO-3b are all in eccentric orbits, but not \corotp. Considering the strong tidal interactions of massive close-in companions with their central star, this suggests an extra body that may have brought -- and maintained -- these planets in their current orbit; something that is not valid for \corotp, and whose formation history might therefore be of different nature. 
\section{Conclusion}

After more than one decade of intensive ground-based extrasolar planet hunting, \corot\ has detected and measured, thanks to the support of ground-based facilities, the fundamental parameters of a massive close-in companion object, located at the overlapping region between the planet and the brown-dwarf domains.  
\corotp\  reopens the debate about the existence of a hitherto non-detected brown-dwarf population at short orbital periods but also about the definition of a planet, such as the common one which, in
this range of mass, relies on the deuterium burning limit.
The exact nature of this new object is therefore still doubtful. Its parameters are in pretty good agreement with the model predictions for brown-dwarfs and if
that is that case, it might simply be the first secure and well-characterized object at the lowest mass end of the stellar population. If  \corotp\  is indeed a brown-dwarf one should review as well the other massive planets like XO-3b, HAT-P2b  and WASP-14b as potential members of this class. An alternative explanation is that \corotp\ belongs to a new, yet unexplored, very massive planet population, widening the variety of exoplanets.  The currently ambiguous nature of \corotp\  makes it therefore a very worthwhile object for further deeper studies. 

\bibliographystyle{aa.bst}
\bibliography{0625}

\begin{thebibliography}{45}
\expandafter\ifx\csname natexlab\endcsname\relax\def\natexlab#1{#1}\fi

\bibitem[{{Aigrain} {et~al.}(2008){Aigrain}, {Collier Cameron}, {Ollivier},
  {Pont}, {Jorda}, {Almenara}, {Alonso}, {Barge}, {Bord{\'e}}, {Bouchy},
  {Deeg}, {de La Reza}, {Deleuil}, {Dvorak}, {Erikson}, {Fridlund}, {Gondoin},
  {Gillon}, {Guillot}, {Hatzes}, {Lammer}, {Lanza}, {L{\'e}ger}, {Llebaria},
  {Magain}, {Mazeh}, {Moutou}, {Paetzold}, {Pinte}, {Queloz}, {Rauer}, {Rouan},
  {Schneider}, {Wuchter}, \& {Zucker}}]{2008A&A...488L..43A}
{Aigrain}, S., {Collier Cameron}, A., {Ollivier}, M., {et~al.} 2008, \aap, 488,
  L43

\bibitem[{{Alibert} {et~al.}(2005){Alibert}, {Mordasini}, {Benz}, \&
  {Winisdoerffer}}]{2005A&A...434..343A}
{Alibert}, Y., {Mordasini}, C., {Benz}, W., \& {Winisdoerffer}, C. 2005, \aap,
  434, 343

\bibitem[{{Alonso} {et~al.}(2008){Alonso}, {Auvergne}, {Baglin}, {Ollivier},
  {Moutou}, {Rouan}, {Deeg}, {Aigrain}, {Almenara}, {Barbieri}, {Barge},
  {Benz}, {Bord{\'e}}, {Bouchy}, {de La Reza}, {Deleuil}, {Dvorak}, {Erikson},
  {Fridlund}, {Gillon}, {Gondoin}, {Guillot}, {Hatzes}, {H{\'e}brard},
  {Kabath}, {Jorda}, {Lammer}, {L{\'e}ger}, {Llebaria}, {Loeillet}, {Magain},
  {Mayor}, {Mazeh}, {P{\"a}tzold}, {Pepe}, {Pont}, {Queloz}, {Rauer},
  {Shporer}, {Schneider}, {Stecklum}, {Udry}, \&
  {Wuchterl}}]{2008A&A...482L..21A}
{Alonso}, R., {Auvergne}, M., {Baglin}, A., {et~al.} 2008, \aap, 482, L21

\bibitem[{{Bakos} {et~al.}(2007){Bakos}, {Shporer}, {P{\'a}l}, {Torres},
  {Kov{\'a}cs}, {Latham}, {Mazeh}, {Ofir}, {Noyes}, {Sasselov}, {Bouchy},
  {Pont}, {Queloz}, {Udry}, {Esquerdo}, {Sip{\H o}cz}, {Kov{\'a}cs},
  {Stefanik}, {L{\'a}z{\'a}r}, {Papp}, \& {S{\'a}ri}}]{2007ApJ...671L.173B}
{Bakos}, G.~{\'A}., {Shporer}, A., {P{\'a}l}, A., {et~al.} 2007, \apjl, 671,
  L173

\bibitem[{{Baraffe} {et~al.}(2008){Baraffe}, {Chabrier}, \&
  {Barman}}]{2008A&A...482..315B}
{Baraffe}, I., {Chabrier}, G., \& {Barman}, T. 2008, \aap, 482, 315

\bibitem[{{Baraffe} {et~al.}(2003){Baraffe}, {Chabrier}, {Barman}, {Allard}, \&
  {Hauschildt}}]{2003A&A...402..701B}
{Baraffe}, I., {Chabrier}, G., {Barman}, T.~S., {Allard}, F., \& {Hauschildt},
  P.~H. 2003, \aap, 402, 701

\bibitem[{{Barge} {et~al.}(2008){Barge}, {Baglin}, {Auvergne}, {Rauer},
  {L{\'e}ger}, {Schneider}, {Pont}, {Aigrain}, {Almenara}, {Alonso},
  {Barbieri}, {Bord{\'e}}, {Bouchy}, {Deeg}, {La Reza}, {Deleuil}, {Dvorak},
  {Erikson}, {Fridlund}, {Gillon}, {Gondoin}, {Guillot}, {Hatzes}, {Hebrard},
  {Jorda}, {Kabath}, {Lammer}, {Llebaria}, {Loeillet}, {Magain}, {Mazeh},
  {Moutou}, {Ollivier}, {P{\"a}tzold}, {Queloz}, {Rouan}, {Shporer}, \&
  {Wuchterl}}]{2008A&A...482L..17B}
{Barge}, P., {Baglin}, A., {Auvergne}, M., {et~al.} 2008, \aap, 482, L17

\bibitem[{{Bouchy} {et~al.}(2008){Bouchy}, {Queloz}, {Deleuil}, {Loeillet},
  {Hatzes}, {Aigrain}, {Alonso}, {Auvergne}, {Baglin}, {Barge}, {Benz},
  {Bord{\'e}}, {Deeg}, {de La Reza}, {Dvorak}, {Erikson}, {Fridlund},
  {Gondoin}, {Guillot}, {H{\'e}brard}, {Jorda}, {Lammer}, {L{\'e}ger},
  {Llebaria}, {Magain}, {Mayor}, {Moutou}, {Ollivier}, {P{\"a}tzold}, {Pepe},
  {Pont}, {Rauer}, {Rouan}, {Schneider}, {Triaud}, {Udry}, \&
  {Wuchterl}}]{2008A&A...482L..25B}
{Bouchy}, F., {Queloz}, D., {Deleuil}, M., {et~al.} 2008, \aap, 482, L25

\bibitem[{{Bouchy} \& {The Sophie Team}(2006)}]{2006tafp.conf..319B}
{Bouchy}, F. \& {The Sophie Team}. 2006, in Tenth Anniversary of 51 Peg-b:
  Status of and prospects for hot Jupiter studies, ed. L.~{Arnold},
  F.~{Bouchy}, \& C.~{Moutou}, 319--325

\bibitem[{{Brown} {et~al.}(2001){Brown}, {Charbonneau}, {Gilliland}, {Noyes},
  \& {Burrows}}]{2001ApJ...552..699B}
{Brown}, T.~M., {Charbonneau}, D., {Gilliland}, R.~L., {Noyes}, R.~W., \&
  {Burrows}, A. 2001, \apj, 552, 699

\bibitem[{{Bruntt} {et~al.}(2002){Bruntt}, {Catala}, {Garrido},
  {Rodr{\'{\i}}guez}, {St{\"u}tz}, {Knoglinger}, {Mittermayer}, {Bouret},
  {Hua}, {Ligni{\`e}res}, {Charpinet}, {Van't Veer-Menneret}, \&
  {Ballereau}}]{2002A&A...389..345B}
{Bruntt}, H., {Catala}, C., {Garrido}, R., {et~al.} 2002, \aap, 389, 345

\bibitem[{{Bruntt} {et~al.}(2008){Bruntt}, {De Cat}, \& {Aerts}}]{bruntt08}
{Bruntt}, H., {De Cat}, P., \& {Aerts}, C. 2008, \aap, 478, 487

\bibitem[{{Claret}(2003)}]{2003A&A...401..657C}
{Claret}, A. 2003, \aap, 401, 657

\bibitem[{{Claret}(2004)}]{2004A&A...428.1001C}
{Claret}, A. 2004, \aap, 428, 1001

\bibitem[{{Deleuil} {et~al.}(2008){Deleuil}, {Meunier}, {Moutou}, {Surace}, {Almenara},
  {Barbieri}, {Debosscher}, {Deeg}, {Granet}, \&
  {Guterman}}]{2008AJ}
{Deleuil}, M., {Meunier}, J.C., {Moutou}, C., {et~al.} 2008, \aj, {\it submitted}

\bibitem[{{Dobbs-Dixon} {et~al.}(2004){Dobbs-Dixon}, {Lin}, \&
  {Mardling}}]{2004ApJ...610..464D}
{Dobbs-Dixon}, I., {Lin}, D.~N.~C., \& {Mardling}, R.~A. 2004, \apj, 610, 464

\bibitem[{{Donati} {et~al.}(2008){Donati}, {Moutou}, {Far{\`e}s}, {Bohlender},
  {Catala}, {Deleuil}, {Shkolnik}, {Cameron}, {Jardine}, \&
  {Walker}}]{2008MNRAS.385.1179D}
{Donati}, J.-F., {Moutou}, C., {Far{\`e}s}, R., {et~al.} 2008, \mnras, 385,
  1179

\bibitem[{{Galland} {et~al.}(2005){Galland}, {Lagrange}, {Udry}, {Chelli},
  {Pepe}, {Queloz}, {Beuzit}, \& {Mayor}}]{2005A&A...443..337G}
{Galland}, F., {Lagrange}, A.-M., {Udry}, S., {et~al.} 2005, \aap, 443, 337

\bibitem[{{Gim{\'e}nez}(2006)}]{2006A&A...450.1231G}
{Gim{\'e}nez}, A. 2006, \aap, 450, 1231

\bibitem[{{Grether} \& {Lineweaver}(2006)}]{2006ApJ...640.1051G}
{Grether}, D. \& {Lineweaver}, C.~H. 2006, \apj, 640, 1051

\bibitem[{{Halbwachs} {et~al.}(2000){Halbwachs}, {Arenou}, {Mayor}, {Udry}, \&
  {Queloz}}]{2000A&A...355..581H}
{Halbwachs}, J.~L., {Arenou}, F., {Mayor}, M., {Udry}, S., \& {Queloz}, D.
  2000, \aap, 355, 581

\bibitem[{{Hinkle} {et~al.}(2000){Hinkle}, {Wallace}, {Valenti}, \&
  {Harmer}}]{hinkle00}
{Hinkle}, K., {Wallace}, L., {Valenti}, J., \& {Harmer}, D. 2000, {Visible and
  Near Infrared Atlas of the Arcturus Spectrum 3727--9300~\AA} (Visible and
  Near Infrared Atlas of the Arcturus Spectrum 3727--9300~\AA\ ed.~Kenneth
  Hinkle, Lloyd Wallace, Jeff Valenti, and Dianne Harmer.~(San Francisco: ASP)
  ISBN: 1-58381-037-4, 2000.)

\bibitem[{{Jackson} {et~al.}(2008){Jackson}, {Greenberg}, \&
  {Barnes}}]{2008ApJ...681.1631J}
{Jackson}, B., {Greenberg}, R., \& {Barnes}, R. 2008, \apj, 681, 1631

\bibitem[{{Johns-Krull} {et~al.}(2008){Johns-Krull}, {McCullough}, {Burke},
  {Valenti}, {Janes}, {Heasley}, {Prato}, {Bissinger}, {Fleenor}, {Foote},
  {Garcia-Melendo}, {Gary}, {Howell}, {Mallia}, {Masi}, \&
  {Vanmunster}}]{2008ApJ...677..657J}
{Johns-Krull}, C.~M., {McCullough}, P.~R., {Burke}, C.~J., {et~al.} 2008, \apj,
  677, 657

\bibitem[{{Joshi} {et~al.}(2008){Joshi}, {Pollacco}, {Collier Cameron},
  {Skillen}, {Simpson}, {Steele}, {Street}, {Stempels}, {Bouchy}, {Christian},
  {Gibson}, {Hebb}, {Hebrard}, {Keenan}, {Loeillet}, {Meaburn}, {Moutou},
  {Smalley}, {Todd}, {West}, {Anderson}, {Bentley}, {Enoch}, {Haswell},
  {Hellier}, {Horne}, {Irwin}, {Lister}, {McDonald}, {Maxted}, {Mayor},
  {Norton}, {Parley}, {Perrier}, {Pont}, {Queloz}, {Ryans}, {Smith}, {Udry},
  {Wheatley}, \& {Wilson}}]{2008arXiv0806.1478J}
{Joshi}, Y.~C., {Pollacco}, D., {Collier Cameron}, A., {et~al.} 2008, ArXiv
  e-prints, 806

\bibitem[{{Knutson} {et~al.}(2007){Knutson}, {Charbonneau}, {Noyes}, {Brown},
  \& {Gilliland}}]{2007ApJ...655..564K}
{Knutson}, H.~A., {Charbonneau}, D., {Noyes}, R.~W., {Brown}, T.~M., \&
  {Gilliland}, R.~L. 2007, \apj, 655, 564

\bibitem[{{Mordasini} {et~al.}(2007){Mordasini}, {Alibert}, {Benz}, \&
  {Naef}}]{Mordasini07}
{Mordasini}, C., {Alibert}, Y., {Benz}, W., \& {Naef}, D. 2007, ArXiv e-prints,
  710

\bibitem[{{Morel} \& {Lebreton}(2007)}]{2007Ap&SS.tmp..460M}
{Morel}, P. \& {Lebreton}, Y. 2007, \apss, 460

\bibitem[{{Murdoch} {et~al.}(1993){Murdoch}, {Hearnshaw}, \&
  {Clark}}]{1993ApJ...413..349M}
{Murdoch}, K.~A., {Hearnshaw}, J.~B., \& {Clark}, M. 1993, \apj, 413, 349

\bibitem[{{Nordstr{\"o}m} {et~al.}(2004){Nordstr{\"o}m}, {Mayor}, {Andersen},
  {Holmberg}, {Pont}, {J{\o}rgensen}, {Olsen}, {Udry}, \&
  {Mowlavi}}]{2004A&A...418..989N}
{Nordstr{\"o}m}, B., {Mayor}, M., {Andersen}, J., {et~al.} 2004, \aap, 418, 989

\bibitem[{{Pinheiro da Silva} {et~al.}(2008){Pinheiro da Silva}, {Rolland},
  {Lapeyrere}, \& {Auvergne}}]{2008MNRAS.384.1337P}
{Pinheiro da Silva}, L., {Rolland}, G., {Lapeyrere}, V., \& {Auvergne}, M.
  2008, \mnras, 384, 1337

\bibitem[{{Queloz} {et~al.}(2001){Queloz}, {Henry}, {Sivan}, {Baliunas},
  {Beuzit}, {Donahue}, {Mayor}, {Naef}, {Perrier}, \&
  {Udry}}]{2001A&A...379..279Q}
{Queloz}, D., {Henry}, G.~W., {Sivan}, J.~P., {et~al.} 2001, \aap, 379, 279

\bibitem[{{Quentin} {et~al.}(2006){Quentin}, {Barge}, {Cautain}, {Meunier},
  {Moutou}, \& {Savalle}}]{Quentin06}
{Quentin}, C., {Barge}, P., {Cautain}, R., {et~al.} 2006, in ESA Special
  Publication, Vol. 1306, ESA Special Publication, 409--415

\bibitem[{{Rentzsch-Holm}(1996)}]{holm96}
{Rentzsch-Holm}, I. 1996, \aap, 312, 966

\bibitem[{{Rieke} \& {Lebofsky}(1985)}]{1985ApJ...288..618R}
{Rieke}, G.~H. \& {Lebofsky}, M.~J. 1985, \apj, 288, 618

\bibitem[{{Schlegel} {et~al.}(1998){Schlegel}, {Finkbeiner}, \&
  {Davis}}]{1998ApJ...500..525S}
{Schlegel}, D.~J., {Finkbeiner}, D.~P., \& {Davis}, M. 1998, \apj, 500, 525

\bibitem[{{Siess}(2006)}]{2006A&A...448..717S}
{Siess}, L. 2006, \aap, 448, 717

\bibitem[{{Southworth}(2008)}]{2008MNRAS.386.1644S}
{Southworth}, J. 2008, \mnras, 386, 1644

\bibitem[{{Surace} {et~al.}(2008) {Alonso}, {Barge}, {Cautain}, {Chabaud}, {Deleuil}, {Fenouillet}, {Meunier}, \& {Moutou}}]{Surace2008}
{Surace}, C., {Alonso}, R., {Barge}, P., {et~al.} 2008, SPIE, Vol. 7019, 132

\bibitem[{{Udry} {et~al.}(2002){Udry}, {Mayor}, {Naef}, {Pepe}, {Queloz},
  {Santos}, \& {Burnet}}]{2002A&A...390..267U}
{Udry}, S., {Mayor}, M., {Naef}, D., {et~al.} 2002, \aap, 390, 267

\bibitem[{{Udry} {et~al.}(2000){Udry}, {Mayor}, {Naef}, {Pepe}, {Queloz},
  {Santos}, {Burnet}, {Confino}, \& {Melo}}]{2000A&A...356..590U}
{Udry}, S., {Mayor}, M., {Naef}, D., {et~al.} 2000, \aap, 356, 590

\bibitem[{{Udry} \& {Santos}(2007)}]{2007ARA&A..45..397U}
{Udry}, S. \& {Santos}, N.~C. 2007, \araa, 45, 397

\bibitem[{{Valenti} \& {Fischer}(2005)}]{2005yCat..21590141V}
{Valenti}, J.~A. \& {Fischer}, D.~A. 2005, VizieR Online Data Catalog, 215,
  90141

\bibitem[{{Valenti} \& {Piskunov}(1996)}]{1996A&AS..118..595V}
{Valenti}, J.~A. \& {Piskunov}, N. 1996, \aaps, 118, 595

\bibitem[{{Wilson} {et~al.}(2008){Wilson}, {Gillon}, {Hellier}, {Maxted},
  {Pepe}, {Queloz}, {Anderson}, {Collier Cameron}, {Smalley}, {Lister},
  {Bentley}, {Blecha}, {Christian}, {Enoch}, {Haswell}, {Hebb}, {Horne},
  {Irwin}, {Joshi}, {Kane}, {Marmier}, {Mayor}, {Parley}, {Pollacco}, {Pont},
  {Ryans}, {Segransan}, {Skillen}, {Street}, {Udry}, {West}, \&
  {Wheatley}}]{2008ApJ...675L.113W}
{Wilson}, D.~M., {Gillon}, M., {Hellier}, C., {et~al.} 2008, \apjl, 675, L113

\bibitem[{{Zacharias} {et~al.}(2004){Zacharias}, {Urban}, {Zacharias},
  {Wycoff}, {Hall}, {Monet}, \& {Rafferty}}]{2004AJ....127.3043Z}
{Zacharias}, N., {Urban}, S.~E., {Zacharias}, M.~I., {et~al.} 2004, \aj, 127,
  3043

\bibitem[{{Zucker} {et~al.}(2004){Zucker}, {Mazeh}, {Santos}, {Udry}, \&
  {Mayor}}]{2004A&A...426..695Z}
{Zucker}, S., {Mazeh}, T., {Santos}, N.~C., {Udry}, S., \& {Mayor}, M. 2004,
  \aap, 426, 695

\end{thebibliography}

\begin{acknowledgements}
 HJD and JMA
acknowledge support by grants ESP2004-03855-C03-03 and
ESP2007-65480-C02-02 of the Spanish Education and Science Ministry.
RA acknowledges support by grant  CNES-COROT-070879.
The German CoRoT Team (TLS and Univ. Cologne) acknowledges DLR
grants 50OW0204, 50OW0603, 50QP07011. Some of the data published in this article were acquired with the 
IAC80 telescope operated by the Instituto de Astrof\'\i sica de
Tenerife at the Observatorio del Teide. The building of the input \corot/Exoplanet catalog was made possible thanks to observations collected for years at the Isaac Newton Telescope (INT), operated on the island of La Palma by 
the Isaac Newton group in the Spanish Observatorio del Roque de Los 
Muchachos of the Instituto de Astrofisica de Canarias. The SME 2.1 package was kindly made
available by N. Piskunov and J. Valenti.
\end{acknowledgements}

\end{document}